\documentclass[a4paper,11pt]{article}
\pdfoutput=1 

\usepackage{jcappub} 

\usepackage[T1]{fontenc} 
\usepackage{booktabs}
\usepackage{subcaption}
\usepackage{enumitem} 
\graphicspath{{./plots_paper/}}

\newcommand{\revise}{}

\title{Forecast of CMB TB and EB correlations for AliCPT-1}
\author[a,b]{Jiazheng Dou,}
\author[c]{Shamik Ghosh,}
\author[d]{Larissa Santos,}
\author[a,b]{Wen Zhao}
\affiliation[a]{ Department of Astronomy, University of Science and Technology of China, Chinese Academy of
Sciences, Hefei, Anhui 230026, P.R.China}
\affiliation[b]{School of Astronomy and Space Sciences, University of Science and Technology of China, Hefei 230026, P.R.China}
\affiliation[c]{Lawrence Berkeley National Laboratory,
Berkeley, CA 94720, U.S.A.}
\affiliation[d]
{Center for Gravitation and Cosmology, Yangzhou University, Yangzhou 224009, P.R.China}
\emailAdd{doujzh@mail.ustc.edu.cn, shamik-ghosh@outlook.com, larissa@yzu.edu.cn, wzhao7@ustc.edu.cn}

\abstract{The correlations between T, E modes and B modes in cosmic microwave background (CMB) radiation, which are expected to vanish under parity symmetry, have become a sensitive probe of the new physics beyond the standard model. In this paper, we forecast the estimation of TB and EB cross power spectra using NILC and cILC on \revise{AliCPT-1 simulations together with Planck HFI and WMAP K maps as ancillary data}. We find that, NILC performs better than cILC on measuring TB and EB correlations in light of its lower uncertainties. In terms of the birefringence angle estimation without assuming systematic errors, the combination of CMB TB and EB spectra from NILC cleaned simulations could reach a sensitivity of \revise{$|\beta|<0.058^\circ$ with 2$\sigma$ significance for the first observing season of AliCPT. Tripling the survey duration will improve this sensitivity to $|\beta|<0.041^\circ$.}}


\begin{document}
\maketitle
\flushbottom

\section{Introduction}

The cosmic microwave background (CMB) photons are known to be partially linearly polarized because of the small quadruple of the radiation field before recombination. Two Stokes parameters used to describe the CMB polarization, Q and U, are typically decomposed into gradient and curl modes, known as E and B modes, in order to construct the rotationally invariant fields under coordinate transformation \cite{zaldarriagaAllSkyAnalysisPolarization1997}. The B modes are sourced from the tensor perturbations produced during inflation, which offers a unique way of searching the primordial gravitational waves (PGWs) in the near future \cite{seljakSignatureGravityWaves1997}.

This paper mainly focuses on the TB and EB correlations, which are often used to test the hypothetical parity violation \cite{lueCosmologicalSignatureNew1999, fengSearchingViolationCosmic2006, quadcollaborationParityViolationConstraints2009}. Parity violation has so far only been present in the weak interaction \cite{leeQuestionParityConservation1956, wuExperimentalTestParity1957}, but may also exist in some extensions of the standard model \cite{alexanderChernSimonsModifiedGeneral2009, marshAxionCosmology2016, ferreiraUltraLightDarkMatter2021}. Under inversion of spatial coordinates $\hat{\boldsymbol{n}}\rightarrow-\hat{\boldsymbol{n}}$, the scalar modes T and E transform as $[T/E]_{\ell m}\rightarrow(-1)^\ell [T/E]_{\ell m}$ while the pseudoscalar mode B transforms as $B_{\ell m}\rightarrow(-1)^{\ell+1} B_{\ell m}$ given the spherical harmonics, $Y_{\ell m}$, have the $(-1)^\ell$ parity. Provided that the CMB temperature and polarization fields are parity invariant and statistically isotropic, the correlations between T and B or E and B would vanish since the TB and EB cross power spectra change the sign under parity transformation (i.e., have the odd parity). Therefore, non-zero values of $C_\ell^{TB}$ or $C_\ell^{EB}$ indicate parity violating physics in the early universe or during photon propagation.

The parity violation in CMB could be induced by either the Faraday rotation from primordial magnetic fields (PMFs) \cite{yadavProbingPrimordialMagnetism2012, deCMBFaradayRotation2013}, or the Chern-Simons term coupling an axion-like pseudo-scalar field and photons \cite{carrollLimitsLorentzParityviolating1990}, both manifesting themselves in CMB as the so-called cosmic birefringence. The birefringence effect rotates the CMB linear polarization by an angle $\beta$, producing a non-zero TB signal as $C_\ell^{TB}=\tan(2\beta)C_\ell^{TE}$ and a EB signal as $C_\ell^{EB}=\frac{\tan(4\beta)}{2}(C_\ell^{EE}-C_\ell^{BB})$ where $C_\ell^{XX},X=\{T,E,B\}$ is the observed CMB power spectrum \cite{fengSearchingViolationCosmic2006,zhaoFluctuationsCosmologicalBirefringence2014, zhaoDetectingRelicGravitational2014}. The Faraday rotation produces a frequency-dependent birefringence angle $\beta(\nu)\propto\nu^{-2}$, which is disfavored by the Planck data with nearly frequency-independent cosmic birefringence \cite{eskiltFrequencyDependentConstraintsCosmic2022,eskiltImprovedConstraintsCosmic2022}. The Chern-Simons interactions change the phase velocity of the left- and right-handed circular polarization of photons and causes a frequency-independent $\beta$. The recent constraint from the joint analysis of the WMAP and Planck data has reached $\beta=0.34^\circ\pm0.09^\circ$ (68\% C.L.) \cite{eskiltImprovedConstraintsCosmic2022}. In addition, the parity violation in gravity can polarize the primordial gravitational waves (PGWs), and also induce the cosmic frequency-independent birefringence, which can rotate the CMB E mode caused by PGWs to B mode, and generate the TB and EB power spectra in the large scales (i.e. low-multipole ranges) \cite{lueCosmologicalSignatureNew1999,pgw2}. Detection of this signal can be used to test the parity symmetry of gravity in the high energy scales \cite{pgw2,pgw3,pgw4,pgw5}.  

Many classes of instrumental systematics could also induce the non-zero TB and EB correlations. BICEP/Keck collaboration \cite{collaborationBICEPKeckXVII2023} summaries 11 distortion fields mixing T, Q and U signals along lines of sight, including the gravitational lensing, the cosmic birefringence, beam or gain mismatches introducing leakage from temperature to polarization (T-P) due to the pair differencing of orthogonal detectors, and E-B leakage from errors in calibration of the polarization angle. Among them, the gravitational lensing itself would not induce non-zero TB and EB signals. The beam (or gain) mismatches can be effectively mitigated by two strategies, the deprojection filtering \cite{sheehyDeprojectingBeamSystematics2019}, or the use of a rapidly rotating half-wave plate (HWP) allowing quick modulation of the polarization signal by rotating the polarization angle at 4 times the rotation frequency of the HWP \cite{johnsonMAXIPOLCosmicMicrowave2007}. However, the residual T-P leakage after deprojection, or the additional systematics due to the non-idealities in realistic HWPs must be evaluated through simulations and jackknife tests \cite{patanchonEffectInstrumentalPolarization2023, monelliImpactHalfwavePlate2022, monelliImpactHalfwavePlate2023}. The miscalibration angles caused by errors in polarization angle calibration are degenerate with the cosmic birefringence angle, but both could be estimated simultaneously from the observed EB correlation since the foreground polarization is only rotated by the miscalibration angle \cite{minamiSimultaneousDeterminationCosmic2019,minamiSimultaneousDeterminationCosmic2020}. Furthermore, Planck has detected a significant TB signal of Galactic dust with a TB/TE ratio of approximately 0.1 \cite{planckcollaborationPlanck2018Results2020e}, which is assumed due to a misalignment between local magnetic fields and filaments of hydrogen clouds \cite{huffenbergerPowerSpectraPolarized2020, clarkOriginParityViolation2021}. These effects creating barriers on verifying parity violation have to be taken carefully into account in the future CMB experiments.

In this work, we forecast the sensitivity of AliCPT to the estimation of the TB and EB cross power spectra \revise{for different survey durations}, approximately providing the tolerance of the potential systematics and the detecting ability of new physics like parity violation. The Ali CMB Polarization Telescope (AliCPT) is a ground-based CMB experiment located in Tibet, China, mainly focusing on the measurement of CMB polarization in the northern hemisphere at 95 GHz and 150 GHz \cite{liProbingPrimordialGravitational2019}. This telescope is expected to begin observing in the near future, and one of its main scientific goals is testing the parity violation, in addition to the direct detection of PGWs through the CMB B-mode polarization \cite{ghoshPerformanceForecastsPrimordial2022,chen01}. 

We use the NILC method \cite{delabrouilleFullSkyLow2009} to combine the input multi-channel CMB maps into one map and compute its TB and EB cross power spectra. We also try the constrained ILC (cILC) \cite{remazeillesCMBSZEffect2011, remazeillesPeelingForegroundsConstrained2021}, a more aggressive method in cleaning foregrounds with an expense of increasing the uncertainty, in order to clean the B modes more radically. Our results show that cILC surpasses NILC in obtaining the cleaned BB auto spectrum, but raising large and unnecessary uncertainties on TB and EB spectra.

Our paper is organized as follows. In Section~\ref{sec:sims}, we describe the sky simulations used in our analysis. In Section~\ref{sec:pipeline}, we present the details of our pipeline. In Section~\ref{sec:results}, we show the results of the cleaned maps and the power spectra, and parameterize the uncertainty of the cross power spectra \revise{in terms of the isotropic birefringence angle}. Finally, we conclude in Section~\ref{sec:conclusions}. Additionally, we introduce the standard ILC method and its variants (NILC and cILC) in Appendix~\ref{sec:methods}.

\section{Simulations}
\label{sec:sims}

\subsection{\revise{AliCPT-1 overview}}
\label{sec:ali-ovv}
\revise{The AliCPT site is located on a 5250 m high peak of the Gangdise mountain, at the geographical coordinates of $80^\circ$ E, $32^\circ$ N, whose first-stage telescope is called AliCPT-1. The first observing season of AliCPT-1 \revise{is expected to start in 2025}. The AliCPT-1 telescope with an aperture of 72 cm and a focal plane diameter of 63.6 cm including 4 modules and 6816 detectors is expected to observe an around 4000 square degree sky patch in its first season deep survey \cite{liProbingPrimordialGravitational2019}. It contains two frequency bands centered at 95 and 150 GHz whose bandwidths are 25\% and 15\% of their center frequencies, respectively. The angular resolutions (FWHM) and the map sensitivities are shown in Table~\ref{tab:instr}.}

The simulated data sets comprise the AliCPT-1 simulations and two ancillary data: the Planck HFI four bands as well as the WMAP K band, to improve the efficiency of foreground cleaning.
The realizations include observations of seven frequency bands: the 95 GHz and 150 GHz bands of the AliCPT-1 experiment, four bands of the Planck HFI instruments (100, 143, 217 and 353 GHz), and the WMAP K band (23 GHz). Each sky map consists of the CMB, thermal noise and astrophysical foregrounds. The CMB and foregrounds are smoothed with the instrumental Gaussian beams listed in Table~\ref{tab:instr}, with \texttt{HEALPix} $N_{side}=1024$ pixelization scheme adopted \cite{gorskiHEALPixFrameworkHigh2005}. The miscalibration of instruments and other systematics are not taken into account and would be considered in the future work.

\begin{table}[tbp]
    \centering
    \caption{Summary table of seven channels involved in our simulations, including WMAP K-band, AliCPT 95GHz and 150GHz bands, and four Planck HFI bands. $\sigma_n^{\rm P}$ denotes the map noise level in $\mu$K-arcmin for polarization. Note that the noise levels are computed by fitting the noise power spectra in our sky patch with the white noise spectra, where the Planck's levels are slightly lower than the values from Table~4 of \cite{planckcollaborationPlanck2018Results2020c}. The ideal channel-combined noise level is about 14$\mu$K-arcmin.}
    \begin{tabular}{lc|cc|cccc}
        \bottomrule
        Instruments&WMAP&\multicolumn{2}{c|}{AliCPT}&\multicolumn{4}{c}{Planck HFI} \\
        \hline
        Frequency(GHz)&23&95&150&100&143&217&353 \\
        \hline
        Beam size(arcmin)&52.8&19&11&9.7&7.3&5.0&4.9 \\
        \hline
        $\sigma_n^{\rm P}$($\mu$K-arcmin)&496&18&24&81&66&92&399\\
        \toprule
    \end{tabular}
    \label{tab:instr}
\end{table}

\subsection{Masks}
\label{sec:masks}
\begin{figure}[tbp]
    \centering
    \begin{subfigure}[b]{0.4\textwidth}
        \centering
        \includegraphics[width=\textwidth]{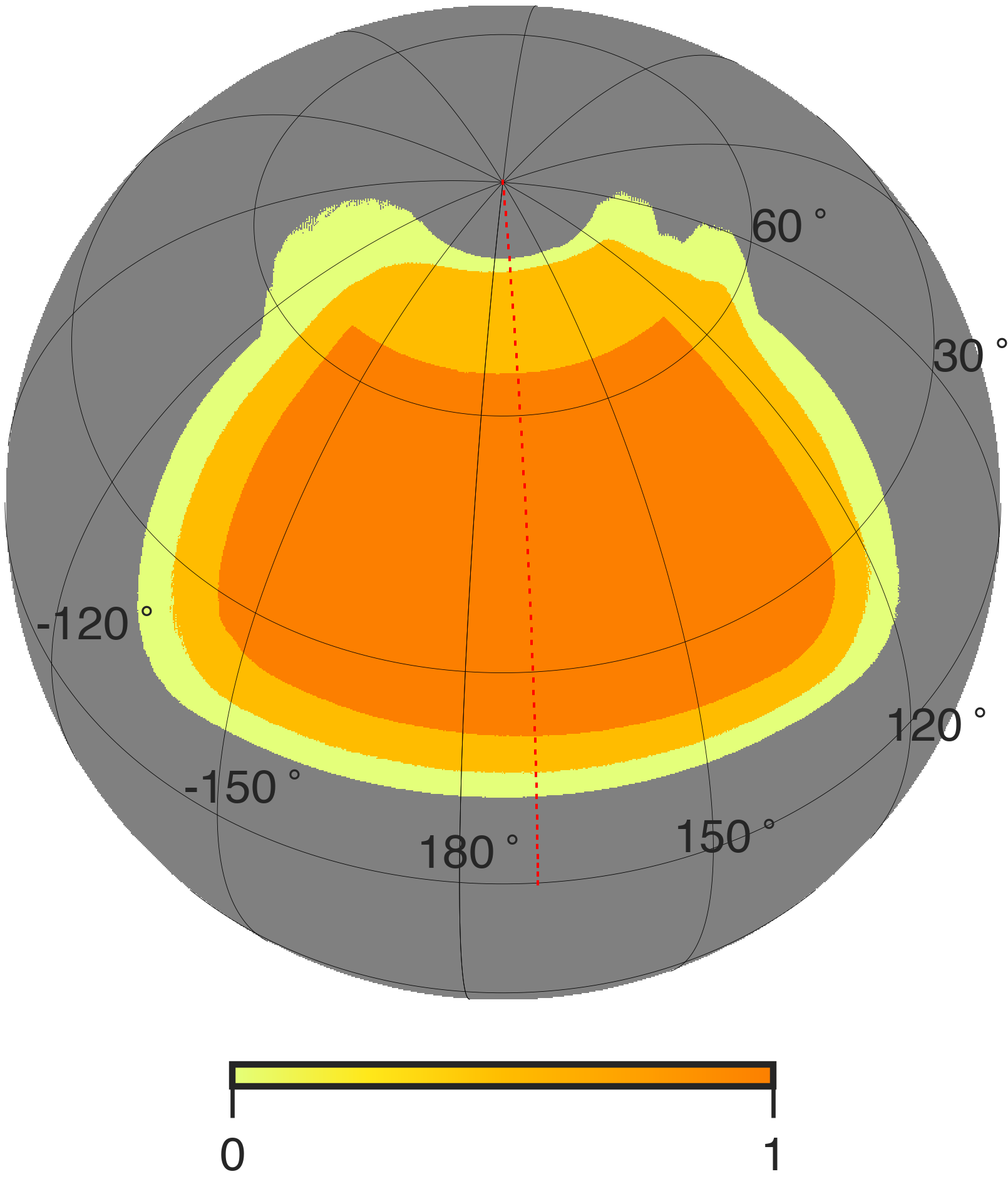}
        \caption{Binary masks}
    \end{subfigure}
    \begin{subfigure}[b]{0.4\textwidth}
        \centering
        \includegraphics[width=\textwidth]{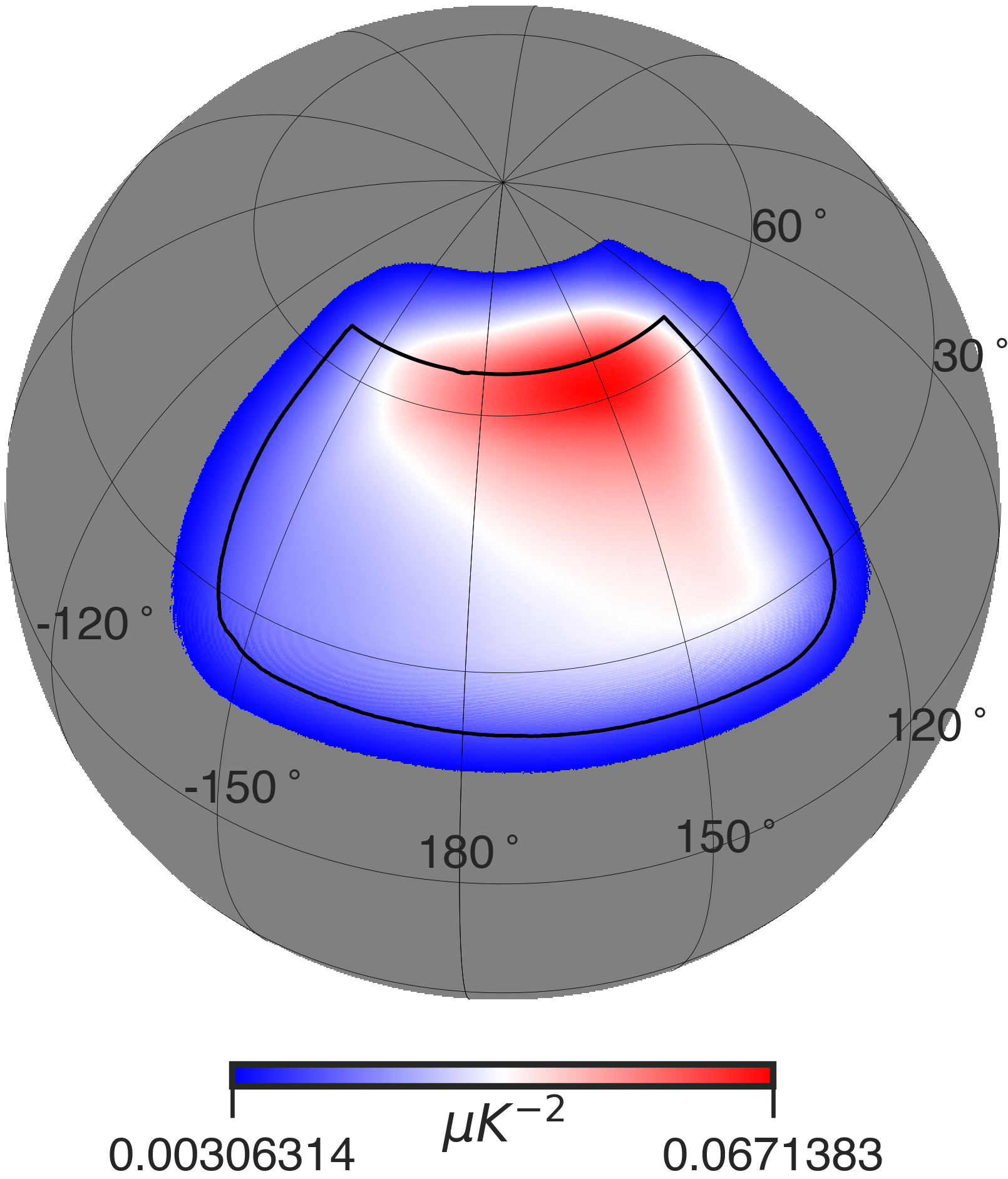}
        \caption{Inverse noise variance}
    \end{subfigure}
    \caption{Masks used in the AliCPT-1 data analysis. (a) From outer to inner are the AliCPT-1 observation patch (yellow), the $20\mu$K mask (light orange), and the UNP mask( dark orange), with sky fractions of 13\%, 10\%, 6.7\%, respectively. \revise{The red dashed line (RA = $171^\circ$) equally divides the $20\mu$K mask into two halves as described in section~\ref{sec:config}.} (b)The inverse variance of the polarization noise at 150 GHz. The black curve shows the boundary of the UNP mask, inside which is the UNP-inv mask.}
    \label{fig:msk}
\end{figure}

Centering at ${\rm RA}=180^\circ$, ${\rm DEC}=30^\circ$ in the northern sky, the AliCPT deep scanning region in the first observing season covers about 17\% of the sky area as shown in Figure~\ref{fig:msk}. Given the non-uniformity of noise and foregrounds in the region, we adopt two binary masks for different cases: for cleaning T and E modes and computing the TE, TB and EB cross power spectra, the `$20\mu$K' mask is applied on T and E maps which masks the pixels with the noise standard deviation above $20\mu$K at \texttt{NSIDE}=1024 for the 150 GHz channel, covering about 10\% of the sky. For cleaning B modes and computing the BB auto-spectrum, we use a smaller mask named the `UNP' mask on B maps, which covers a cleaner region of sky ($f_{\rm sky}=6.7\%$) in terms of both noise and foregrounds, so that the use of the UNP mask introduces less contamination into B modes.
We also define the UNP-inv mask as the UNP mask divided by the variance of the polarization noise at 150 GHz. The masks are shown in Figure~\ref{fig:msk}.

\subsection{CMB}
The CMB maps are Gaussian realizations generated from the power spectra obtained by \texttt{CAMB} \cite{lewisEfficientComputationCMB2000} package using the best-fit Planck 2018 parameters\footnote{Specifically, the parameters are dark matter density $\Omega_ch^2=0.120$, baryon density $\Omega_bh^2=0.02237$, scalar spectral index $n_s=0.9649$, optical depth $\tau=0.0544$, Hubble constant $H_0=69.36\ {\rm km\ s^{-1}\ Mpc^{-1}}$ and the primordial comoving curvature power spectrum amplitude $A_s=2.10\times10^{-9}$.} \cite{aghanimPlanck2018Results2020a}. The tensor-to-scalar ratio is set to be 0.03. Note that the B mode induced by the primordial tensor mode is much smaller than the lensing B mode at scales of interest ($\ell>40$), thus making negligible effects on the uncertainty of the cross power spectra that depend on the lensed B mode power spectrum (see Equation~\ref{eq:sigma-dl}). The lensing effects are added by the \texttt{LensPyx} \cite{reineckeImprovedCMBLensing2023} package. 

\subsection{Noise}
\label{sec:noise}
For AliCPT 95, 150 GHz and WMAP K bands, the instrumental noise simulations are Gaussian white realizations sampled from the noise variance map of AliCPT-1 bands and WMAP K band (upgraded to $N_{\rm side}=1024$), respectively. We assume the possible $1/f$ noise could be effectively suppressed by pair difference and polynomial filtering. For four Planck HFI bands, we use 300 Planck FFP10 noise simulations from the Planck Legacy Archive\footnote{http://pla.esac.esa.int/pla}.

The effective polarization noise levels for seven channels, obtained by fitting the debeamed noise power spectrum at $\ell=[30,360]$ in the AliCPT sky patch with the white noise spectrum, are tabulated in Table~\ref{tab:instr}. The ideal noise level combining the data of all channels is about 10 $\mu$K-arcmin, given by $\mathcal{N}_\ell=[\sum_\nu \mathcal{N}_{\ell,\nu}^{-1}]^{-1}$. The temperature noise levels are $\sqrt{2}$ smaller than polarization. 

\subsection{Foregrounds}
\label{sec:fg}
The foreground simulations are generated by the \revise{Planck Sky Model (PSM \cite{delabrouillePrelaunchPlanckSky2013}) package}. The diffuse foreground components include synchrotron, thermal dust, anomalous microwave emission (AME), free-free and CO emissions, where the synchrotron emission and the thermal dust emission predominate among them. Once we consider the correlations between the scalar modes and B modes, especially the EB power spectrum, the contribution from other foreground components becomes negligible. We do not involve the point sources in our main analysis because they are masked by the point source mask and the remnant effects are weak. In our foreground model, the thermal dust polarization maps are generated from the GNILC template of the Planck 2018 release \cite{planckcollaborationPlanck2018Results2020a}, scaled to different frequencies by a modified blackbody SED with the \revise{spatially-varying} dust temperature and spectral indices from the GNILC Planck 2015 dust maps best-fit \revise{at $5^\prime$ resolution} \cite{planckcollaborationPlanckIntermediateResults2016}. The synchrotron polarization template is based on the Planck 2018 SMICA map, and it is scaled by a power law SED with a fixed $\beta_s$ of -3.08.

The TB correlation of foregrounds, mainly due to the thermal dust emissions, is detected by Planck \cite{planckcollaborationPlanck2018Results2020e}, and the EB correlation of dust is speculated from Planck Data Release 4 given the different birefringence angles estimated with different sky fractions \cite{diego-palazuelosCosmicBirefringencePlanck2022}. The EB spectrum of dust can be modeled as \cite{clarkOriginParityViolation2021, diego-palazuelosCosmicBirefringencePlanck2022}
\begin{equation}\label{eq:eb-dust}
    C_\ell^{EB, \rm dust}=A_\ell C_\ell^{EE, \rm dust}\sin(4\psi_\ell^{\rm dust})\,,
\end{equation}
with the free parameter $0\leq A_\ell \ll 1$ and the angle $\psi_\ell^{\rm dust}=\frac{1}{2}\arctan(C_\ell^{TB, \rm dust}/C_\ell^{TE, \rm dust})$ which can be interpreted as the misalignment angle between the long axis of hydrogen filaments and the interstellar magnetic field orientation. When the misalignment occurs, TB and EB correlations emerge with the same sign. $C_\ell^{TB, \rm dust}$ above is computed from the Planck 353 GHz data.
However, the EB correlation (in terms of Equation~\ref{eq:eb-dust}) is not robustly constrained under Planck's sensitivity, and the positive dust TB signal is not adequately modeled by any Python Sky Model (PySM) from \texttt{d0} to \texttt{d12} \cite{thornePythonSkyModel2017,zoncaPythonSkyModel2021}.
Therefore, we do not consider these dust correlations in our sky model, and we have verified that \revise{all built-in} models in PySM have negligible foreground contributions on TB and EB spectra as the AliCPT model does.


\section{Pipeline}
\label{sec:pipeline}
Our component separation procedure is independently applied on T, E and B modes. We choose NILC, a blind method without any prior about the properties of noise and foregrounds, given the lack of knowledge about polarized foregrounds. It is implemented in needlet domain which divides the data into super-pixels with a localization in both pixel space and harmonic space. More details of this method is introduced in Appendix~\ref{sec:nilc}.

First, we use the \texttt{healpy} package to decompose the input IQU maps to the TEB maps, while for B modes we use the template cleaning method of Liu, et.al \cite{liuMethodsPixelDomain2019} to get rid of the so-called E-B leakage, \revise{which is briefly introduced in Appendix~\ref{sec:tc}. We apply the $20\mu$K mask introduced in section~\ref{sec:masks} onto the TEB maps after template cleaning. } Then we use the NILC method independently on the T, E and B modes after reconvolving the multi-channel maps into a common beam of FWHM=11 arcmin. For B modes, as a compromise for the case of estimating the BB spectrum, we consider the (harmonic) constrained ILC (cILC), i.e. adding more constraints on the ILC weights to null dominant foreground components (see Appendix~\ref{sec:cilc} for details), since the standard NILC leaves considerable foreground residuals on the BB spectra which significantly biases the tensor-to-scalar ratio $r$ in our tests \cite{douForegroundRemovalILC2023}. The cILC in harmonic space and in needlet space are both \revise{capable} to clean the \revise{high Galactic latitude} foregrounds in our sky patch and produce similar results, while the former saves nearly half of computational hours thus being used in our pipeline.\footnote{\revise{The comparison of the performance of cILC in different domains on AliCPT is detailed in \cite{douForegroundRemovalILC2023}.}} Finally we compute the auto and cross spectra of cleaned T, E and B modes by the \texttt{NaMASTER} package\footnote{https://github.com/LSSTDESC/NaMaster} \cite{alonsoUnifiedPseudoC_2019}. Apodization is applied during the power spectrum estimation to avoid high residuals on the boundary of the sky patch. We adopt a 6$^\circ$ C2 apodization of the $20\mu$K mask \revise{except for the case of computing the BB power spectrum where we apply the UNP-inv mask}. For the auto power spectra, we debias the noise power spectrum using the average of 100 noise realizations projected by the ILC weights. We do no noise debiasing for the cross spectra since the residual noise power spectrum is negligible compared to the uncertainty.

The NILC method has two free parameters, the needlet bands controlling the localization in harmonic space, and the ILC bias ratio between the ILC bias and the input CMB power spectrum \cite{delabrouilleFullSkyLow2009}. We set the cosine needlets with peaks at $\ell_{mid}^j=[15,30,60,120,300,700,1200]$ (see Appendix~\ref{sec:nilc} for details) and a 3\% ILC bias which limits the loss of CMB power spectra to 3\% (same for cILC). 

\revise{Since we use no TOD simulation of AliCPT, the TOD filters introduced by \cite{ghoshPerformanceForecastsPrimordial2022} are not involved in this work. However, the measurements of TB and EB correlations may suffer from the side effects of the two AliCPT TOD filters in practical, specifically the power suppression and E-B leakage, which could be calibrated in a similar way as \cite{ghoshPerformanceForecastsPrimordial2022}. The former suppresses the power spectrum by a transfer function dependent on scales, and the latter induces a spurious EB correlation between the E-modes and the E-B leakage in B-modes. The tests on TOD simulations are beyond the scope of this article and will be accomplished in the future work.
}

\section{Results}
\label{sec:results}
Our results consist of the foreground cleaned maps of T, E, B modes presented in section~\ref{sec:maps}, and the auto and cross spectra computed from the cleaned maps after noise debiasing. The TB and EB cross spectra with B-modes cleaned by NILC and cILC methods are compared in section~\ref{sec:ps}. In section~\ref{sec:param} we design a $\chi^2$ test on the TB and EB spectra to parameterize their uncertainties, \revise{and we finally present the results of other instrumental configurations in section~\ref{sec:config}}.

\subsection{Maps}
\label{sec:maps}

\begin{figure}[tbp]
    \centering
    \begin{subfigure}[b]{0.3\textwidth}
        \centering
        \includegraphics[width=\textwidth]{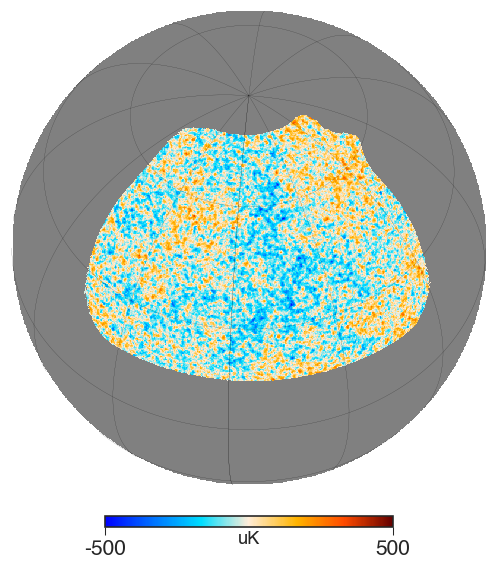}
        \caption{input T CMB}
    \end{subfigure}
    \begin{subfigure}[b]{0.3\textwidth}
        \centering
        \includegraphics[width=\textwidth]{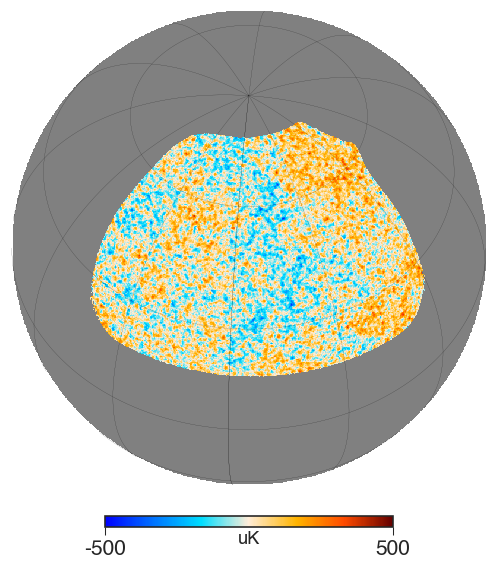}
        \caption{NILC cleaned T map}
    \end{subfigure}
    \begin{subfigure}[b]{0.3\textwidth}
        \centering
        \includegraphics[width=\textwidth]{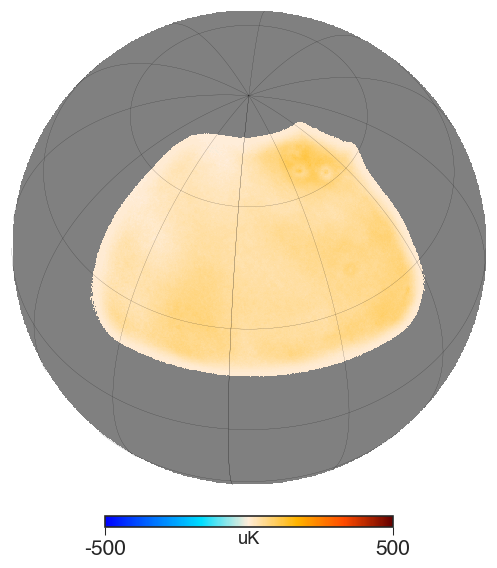}
        \caption{NILC residual T map}
    \end{subfigure}
    \\
    \begin{subfigure}[b]{0.3\textwidth}
        \centering
        \includegraphics[width=\textwidth]{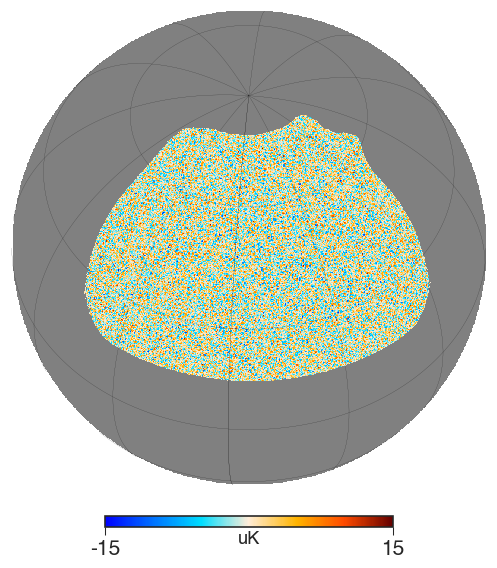}
        \caption{input E CMB}
    \end{subfigure}
    \begin{subfigure}[b]{0.3\textwidth}
        \centering
        \includegraphics[width=\textwidth]{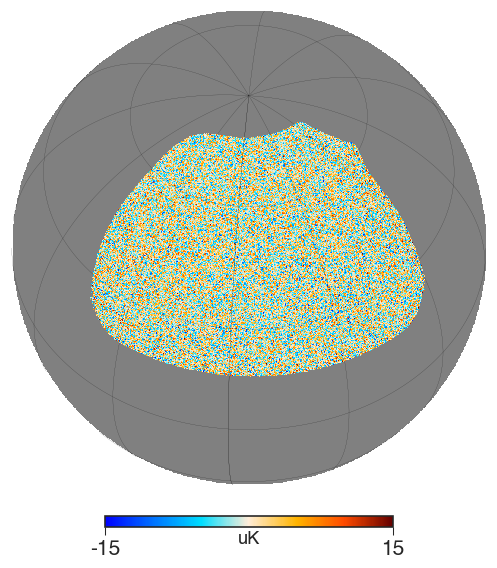}
        \caption{NILC cleaned E map}
    \end{subfigure}
    \begin{subfigure}[b]{0.3\textwidth}
        \centering
        \includegraphics[width=\textwidth]{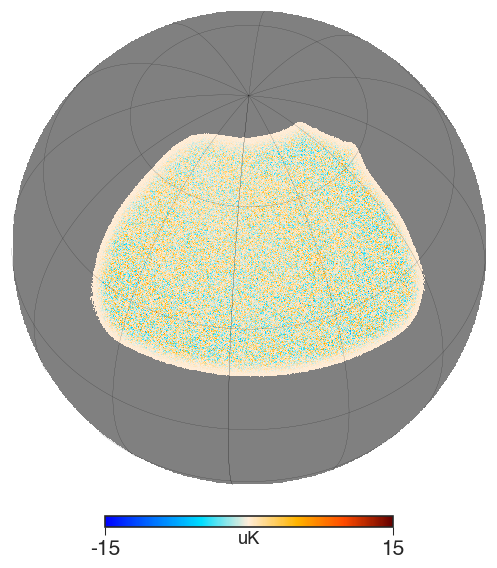}
        \caption{NILC residual E map}
    \end{subfigure}
    \caption{The input CMB maps (left column), the foreground cleaned maps (middle column) and their difference, the residual maps (right column) for T and E modes. The first row is for T modes and the second row for E modes.}
    \label{fig:TE-maps}
\end{figure}

\begin{figure}[tbp]
    \centering
    \begin{subfigure}[b]{0.3\textwidth}
        \centering
        \includegraphics[width=\textwidth]{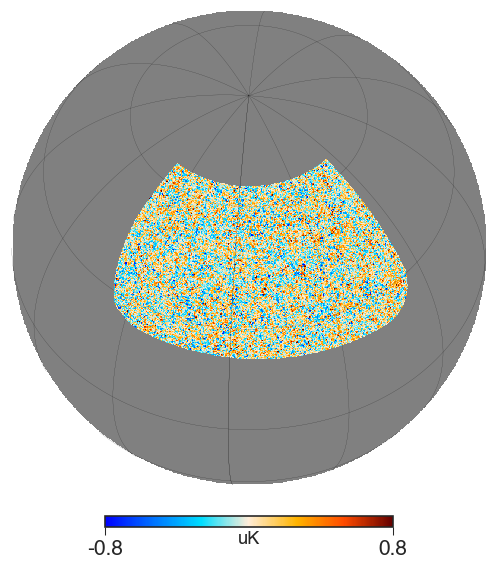}
        \caption{input B CMB}
    \end{subfigure}
    \begin{subfigure}[b]{0.3\textwidth}
        \centering
        \includegraphics[width=\textwidth]{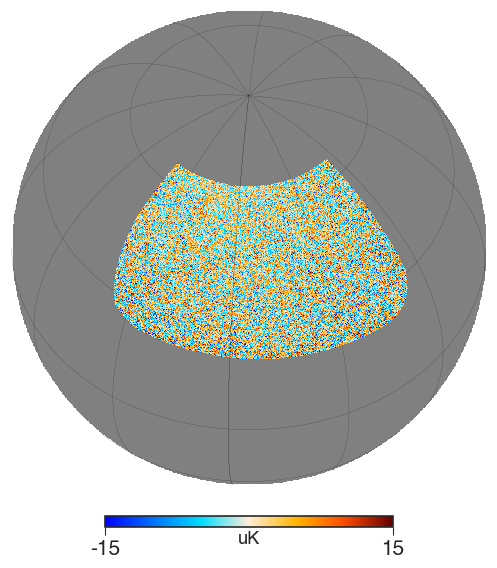}
        \caption{cILC cleaned B map}
    \end{subfigure}
    \begin{subfigure}[b]{0.3\textwidth}
        \centering
        \includegraphics[width=\textwidth]{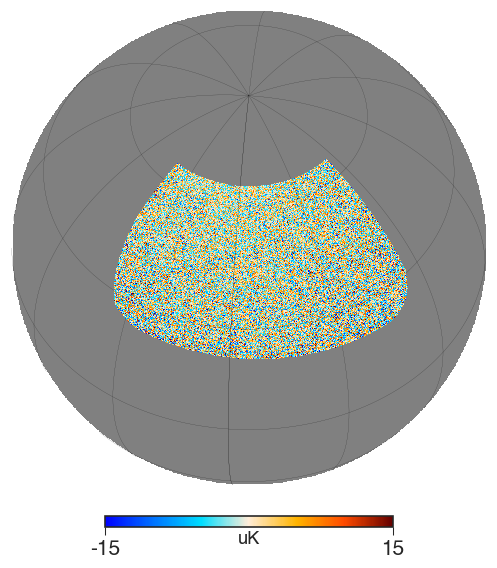}
        \caption{cILC residual B map}
    \end{subfigure}
    \\
    \begin{subfigure}[b]{0.3\textwidth}
        \centering
        \includegraphics[width=\textwidth]{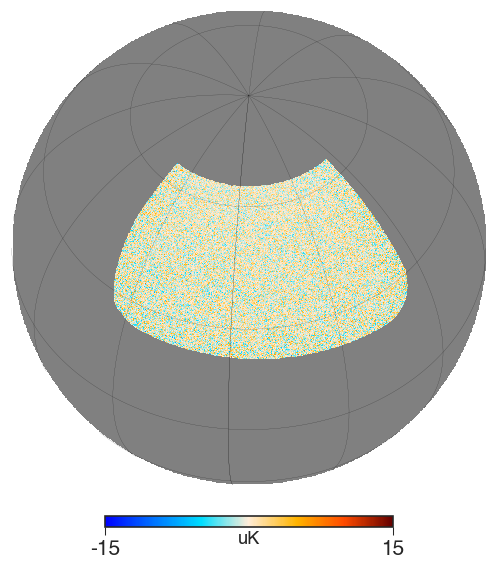}
        \caption{NILC cleaned B map}
    \end{subfigure}
    \begin{subfigure}[b]{0.3\textwidth}
        \centering
        \includegraphics[width=\textwidth]{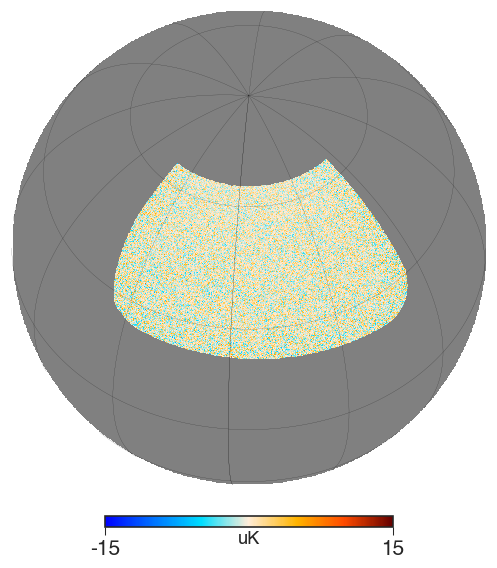}
        \caption{NILC residual B map}
    \end{subfigure}
    \caption{The input CMB map (top left), the foreground cleaned maps, and the residual maps for B modes. Both NILC and cILC are used to clean B modes.}
    \label{fig:B-maps}
\end{figure}

We show the input CMB maps, the foreground cleaned maps, and their differences which are the so-called `residual' maps in Figure~\ref{fig:TE-maps} and Figure~\ref{fig:B-maps}, for T, E modes and B modes, respectively. At map level, the NILC effectively recover the input CMB signal for T and E modes. For B modes, however, the residual is much larger than the CMB signal since the signal-to-noise ratio for B modes is relatively low within our instrumental configuration. Compared to NILC, the cILC renders higher residual (mainly noise residual) on the maps. At the power spectrum level, the noise residual can be properly counteracted by the estimation of the noise power spectrum given the prior knowledge of noise properties, i.e. the noise debiasing procedure.

\subsection{Power spectra}
\label{sec:ps}
\begin{figure}[tbp]
    \centering
    \begin{subfigure}[b]{0.7\textwidth}
        \centering
        \includegraphics[width=\textwidth]{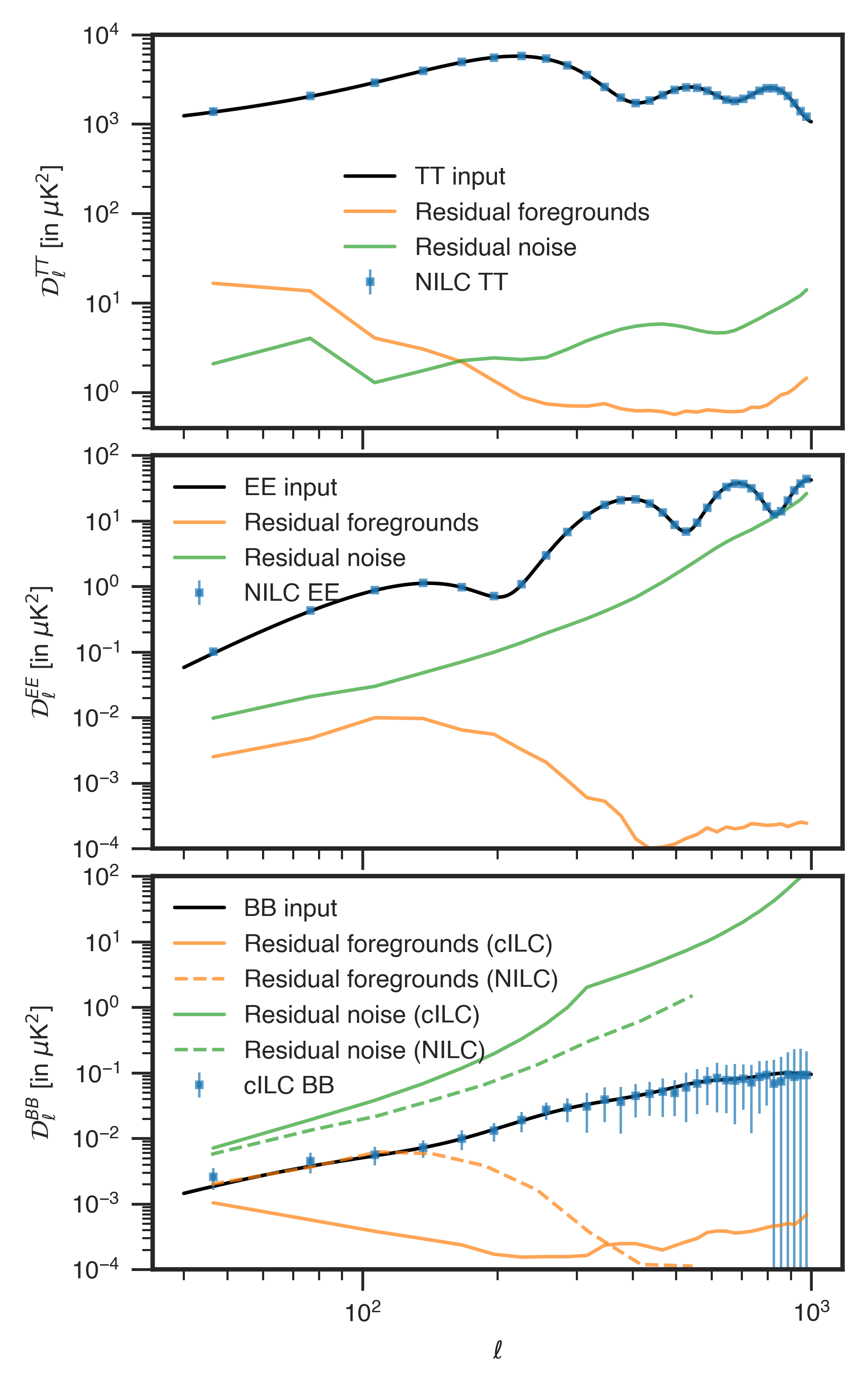}
    \end{subfigure}
    \caption{The auto power spectra including NILC-cleaned TT (top), EE (middle) and cILC-cleaned BB (bottom). The input CMB (black), the reconstructed CMB (blue), the residual noise (green) power spectra are plotted in each panel. For BB, as a comparison, we also plot the noise and foreground residuals for NILC-cleaned B maps (dashed).}
    \label{fig:auto-spec}
\end{figure}

We compute the TT, EE, BB, TE, TB and EB power spectra from the foreground cleaned maps using the pseudo-$C_\ell$ method, with the multipole range $\ell=[30,990]$. The multipole bin size is chosen as $\Delta\ell=30$. The reconstructed CMB power spectra of TT, EE and BB are plotted in Figure~\ref{fig:auto-spec}. The T and E modes are cleaned by NILC, while the B modes are cleaned by harmonic cILC. The input power spectra are well recovered by the foreground cleaning pipeline, given that the residual foregrounds are much lower than the CMB signal, especially at small scales about $\ell>200$. As a comparison, we also plot the BB noise and foreground residuals for NILC-cleaned B maps in the bottom panel. As for the BB spectrum, cILC outperforms NILC on reducing the residual foregrounds but, in exchange, leaving higher noise residual which can be properly debiased given the prior knowledge of noise properties, though. 

\begin{figure}[tbp]
    \centering
    \begin{subfigure}[b]{0.7\textwidth}
        \centering
        \includegraphics[width=\textwidth]{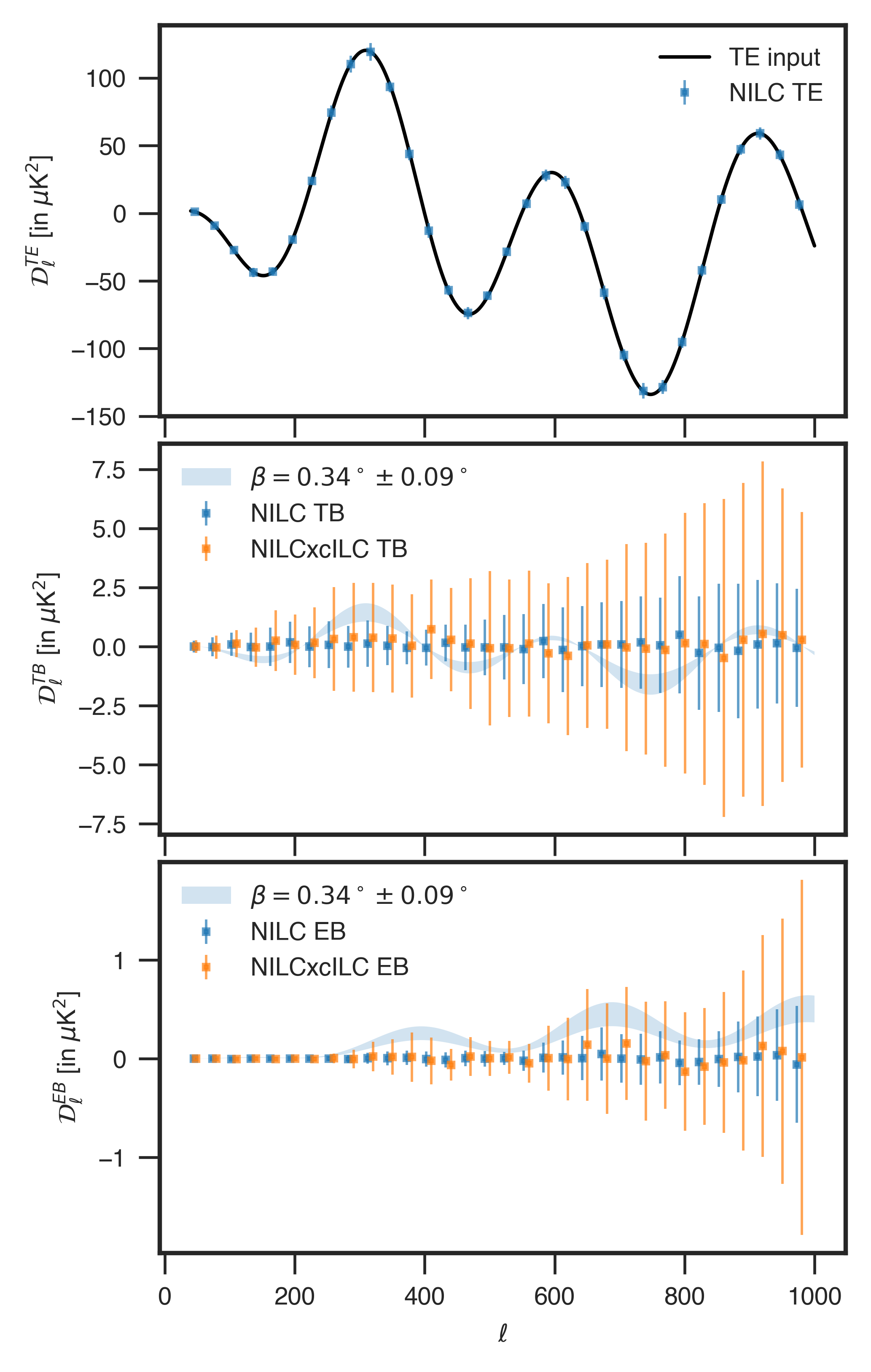}
    \end{subfigure}
    \caption{The cross power spectra of TE (top), TB (middle) and EB (bottom). The NILC-cleaned spectra (blue) and the spectra from NILC-cleaned T or E modes crossing the cILC-cleaned B modes (orange) are compared in the TB and EB plots. The TB and EB spectra induced by the cosmic birefringence angle of $\beta=0.34^\circ\pm0.09^\circ$, the current best constraint of $\beta$ \revise{from WMAP+Planck data }\cite{eskiltImprovedConstraintsCosmic2022}, are added as the shaded areas.}
    \label{fig:cross-spec}
\end{figure}

The reconstructed power spectra of TE, TB and EB are plotted in Figure~\ref{fig:cross-spec}. The T and E modes are cleaned by NILC, while the B modes are either cleaned by NILC or harmonic cILC. For the TB and EB cross spectra, the noise and foreground correlations between the scalar modes and the tensor mode almost vanish. The intrinsic TB correlations from thermal dust emissions are not modeled in our baseline foregrounds, as referred in section~\ref{sec:fg}. The TB and EB spectra of both NILC and NILC$\times$cILC are consistent with zero, implying that the empirical correlations by chance, \revise{i.e. the non-zero values of the mean cross spectra, are small compared to their uncertainties}. Therefore, the low uncertainty of NILC makes it a more appropriate method of estimating TB and EB correlations than cILC.

\subsection{Parameterization of sensitivity}
\label{sec:param}
To parameterize the uncertainty of the TB and EB correlations, we attribute the possible non-zero TB and EB signals to the cosmic birefringence\footnote{The miscalibration of the polarization angle is completely degenerate with the birefringence in the NILC power spectra. Therefore, we plan to firstly estimate birefringence assuming no systematics, and then compare its sensitivity with the forecasted uncertainty of the polarization angle calibration, which depends on the calibration method and the instrumental errors. For instance, the approach taking the Crab nebula (or Tau A) as the reference source is limited by the lowest uncertainty of the polarization observations of Tau A, which is 0.27$^\circ$ to date \cite{aumontAbsoluteCalibrationPolarisation2020}.}\footnote{Note that, in the real data analysis, the effect of cosmic birefringence should be separated from other contaminations, including various systematics, foreground radiations and so on \cite{minamiNewExtractionCosmic2020,diego-palazuelosCosmicBirefringencePlanck2022,eskiltImprovedConstraintsCosmic2022,eskiltCosmoglobeDR1Results2023}. However, in this article, we will not analyze these details, but focus on the construction of the clean TB and EB spectra. }.
Assume that the observed CMB power spectra are rotated by a birefringence angle of $\beta$, and the null hypothesis that the parity symmetry is preserved. The TB and EB observed spectra are given by $C_\ell^{TB}(\beta)=\tan(2\beta)C_\ell^{TE}$ and $C_\ell^{EB}(\beta)=\frac{1}{2}\tan(4\beta)(C_\ell^{EE}-C_\ell^{BB})$, respectively. Our aim is to estimate $\beta$ from the null ($\beta=0$) simulations and find the 95\% threshold of their distribution, so that the data with estimated $\beta$ above which would reject the null hypothesis with a 95\% confidence level. We use the $\chi^2$ statistics defined as
\begin{subequations}
    \begin{align}
    \chi^2_{X,b}(\beta) &=\sum_{b'}[C_b^X(\beta)-C_{b,n}^X] [M_{bb'}^X]^{-1} [C_{b'}^X(\beta)-C_{b',n}^X]\,,
    \\
    \chi^2_X(\beta) &=\sum_{bb'}[C_b^X(\beta)-C_{b,n}^X] [M_{bb'}^X]^{-1} [C_{b'}^X(\beta)-C_{b',n}^X]\,,\label{eq:chi2}
    \end{align}
\end{subequations}
where $X=TB$ or $EB$, $b$ denotes the multipole bin, $N_b$ is the number of bins, $C_{b,n}^X$ are null simulations with no input birefringence and $M_{bb'}^X=\langle C_{b,n}^X C_{b',n}^X\rangle$. The $\chi^2$ statistics characterize the statistical consistency between the model with $\beta$ and the null simulations, where $\chi^2_{X,b}$ estimates at each multipole bin and $\chi^2_X=\sum_{b}\chi^2_{X,b}$ summaries over a range of multipole bins. We can estimate $\beta$ for each null simulation by minimizing $\chi^2_X(\beta)$. In addition, we could estimate a joint $\beta$ combining TB and EB by minimizing $\chi^2(\beta)=\chi^2_{TB}(\beta)+\chi^2_{EB}(\beta)$.

\begin{figure}[tbp]
    \centering
    \begin{subfigure}[b]{0.59\textwidth}
        \centering
        \includegraphics[width=\textwidth]{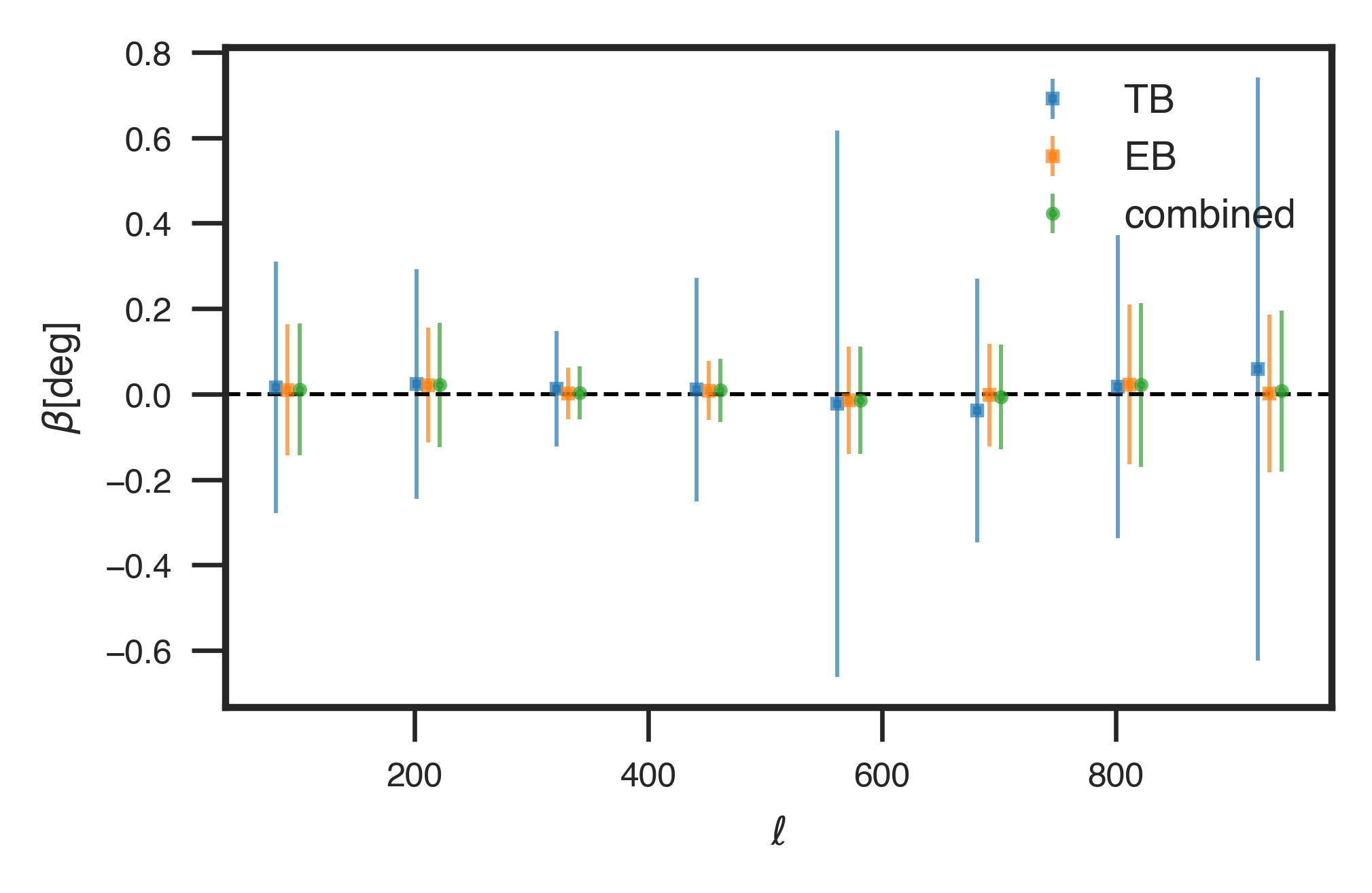}
    \end{subfigure}
    \begin{subfigure}[b]{0.4\textwidth}
        \centering
        \includegraphics[width=\textwidth]{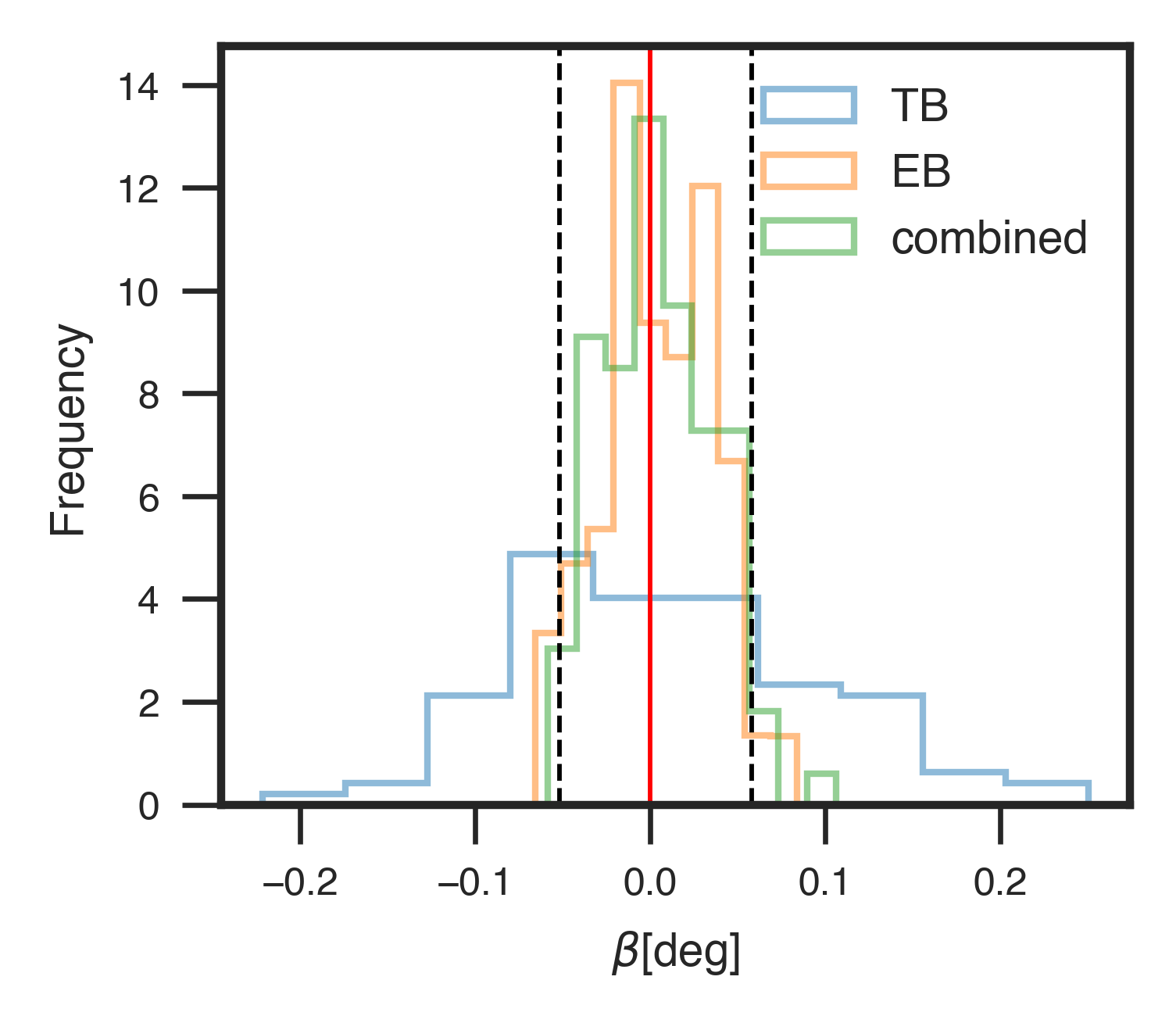}
    \end{subfigure}
    \caption{Left: the best-fit $\beta$ minimizing $\chi^2_{X,b}$ and their 1$\sigma$ uncertainties versus multipole with the range $\ell=[40,1000]$. Right: the distribution of the $\beta$ estimated from 100 null simulations. The 95\% C.L. boundaries for the TB-EB combined case are plotted as the black dashed vertical lines. The input value 0 is labelled as the red line.}
    \label{fig:chi2}
\end{figure}

The left panel of Figure~\ref{fig:chi2} plots the $\beta$ values minimizing $\chi^2_{X,b}$ against multipoles, computed from the NILC-cleaned TB and EB cross spectra. Note that the uncertainties of EB are smaller than that of TB at all scales, indicating that the EB spectrum is a more sensitive probe of parity violation than the TB spectrum.\footnote{The essential reason is that the uncertainty of TB is larger than that of EB, given the cosmic variance of T-mode much larger than that of E-mode.} As expected, the TB-EB combined case has the similar uncertainties with the EB-only case, since the combination is dominated by information from EB. The distribution of $\beta$ minimizing $\chi^2_{X}$ from 100 null simulations is plotted in the right panel of Figure~\ref{fig:chi2}. The mean values, 1$\sigma$ and 2$\sigma$ uncertainties of $\beta$ are tabulated in Table~\ref{tab:beta*}. The best-fit $\beta$ for the TB-EB combined case is 0.005$\pm$0.032 deg. 

\revise{Due to the limit number of simulations, the non-zero mean value of null simulations is of order $\sigma(\beta)/\sqrt{N_{\rm sim}}$ where $N_{\rm sim}=100$, and the distribution of $\beta$ is not perfectly Gaussian as shown in the right panel of Figure~\ref{fig:chi2}. Both effects lead to the asymmetry of the 1$\sigma$ and 2$\sigma$ confidence intervals about zero. To avoid confusion, hereafter we conservatively take the larger abstract of the upper and lower bounds of the 95\% confidence interval of $\beta$ to be the 2$\sigma$ uncertainty as if we have enough simulations. From the combined analysis of TB and EB correlations, we conclude that the null hypothesis ($\beta=0$) could be rejected with 2$\sigma$ significance given the data with a birefringence angle with absolute value greater than 0.058$^\circ$.}

\revise{In \cite{liProbingPrimordialGravitational2019} the authors have used the MCMC sampler to estimate the constraint of $\beta$ and concluded that after 3 years' observation with annually increasing module numbers (24 module*year, as referred to in Table 1 of \cite{liProbingPrimordialGravitational2019}), the AliCPT is capable of providing a constraint on the average rotation angle of $\sigma(\beta)\sim0.01^\circ$. Under the first year's observation (4 module*year adopted in this work), their extrapolated uncertainty should be $\sigma(\beta)\sim0.024^\circ$.\footnote{Roughly speaking, $\sigma(\beta)\propto \sigma(\hat{\mathcal{D}}_\ell^{EB})\propto \sqrt{\mathcal{N}_\ell^{P}}\propto\sigma_n^P$.} Since they consider no foreground but the optimal sensitivity of AliCPT, this result is slightly better than our uncertainty of $\sigma(\beta)=0.032^\circ$.}

Given that the uncertainty of current calibration approaches is well beyond this constraint, we caution that this constraint might be too optimistic. Therefore, we do not recommend directly estimating the birefringence angle without considering systematics in the real analysis.

\begin{table}[tbp]
    \centering
    \caption{The mean estimated $\beta$, its 1$\sigma$ and 2$\sigma$ uncertainties from the NILC-cleaned TB and EB spectra of 100 null simulations.}
    \begin{tabular}{lcc}
        \bottomrule
        Case & $\beta$[deg] & 95\% C.L. of $\beta$[deg] \\
        \hline
        TB & 0.011$\pm$0.087 & (-0.134, 0.186) \\
        \hline
        EB & 0.004$\pm$0.032 & (-0.056, 0.057) \\
        \hline
        combined & 0.005$\pm$0.032 & (-0.052, 0.058) \\
        \toprule
    \end{tabular}
    \label{tab:beta*}
\end{table}

\subsection{\revise{Impacts of varying the instrumental configurations}}
\label{sec:config}

The motivation for investigating the impacts of new instrumental configurations, such as the sky coverage, the map depth or sensitivity and the integration time, is not only to forecast on longer survey duration, but also based on two facts: the practical observational performance might not reach our optimistic objectives given some unpredictable systematics and detector breakdowns; the optimization of the configuration parameters in terms of TB/EB correlations would provide insights to the design of the future scanning strategy of AliCPT or other ground-based CMB experiments.

In this section we aim to modify the configuration parameters including the map sensitivity of AliCPT, the sky coverage and the integration time (which depends on the previous two parameters), and evaluate their impacts on the sensitivity to TB or EB correlations. Specifically, besides the baseline scheme, we have tried these scenarios: (1) we modify the polarization map-depth of AliCPT bands by a factor of $\sqrt{2}$, $1/\sqrt{2}$, $1/\sqrt{3}$, corresponding to an integration time of 2, 8, 12 module*year respectively; (2) we keep the integration time unchanged (4 module*year) but increasing the \revise{map-depth} by $\sqrt{2}$ while cutting a half of the sky coverage to mimic an alternative scanning strategy. We cut the 20$\mu$K mask in two halves along the equatorial longitude line of $171^\circ$, whose noise levels are nearly the same. The left half covering 5\% of the sky is then adopted as the mask in this scenario. We propose to test five scenarios in total, whose detailed configuration parameters are listed in Table~\ref{tab:config}. The beam-combined noise level $\sigma_n^P$ is obtained by fitting the inverse weighted noise power spectrum $\mathcal{N}_\ell=[\sum_\nu \mathcal{N}_{\ell,\nu}^{-1}]^{-1}$ with the white noise power spectrum $\mathcal{N}_\ell^{wh}=\frac{\ell(\ell+1)}{2\pi}(\pi\sigma_n^P/10800)^2$ at $\ell=[30,360]$.

\begin{table}[tbp]
    \centering
    \caption{The modified configuration parameters (including the sky coverage, the beam-combined noise level and the integration time) of five scenarios, their uncertainties of the NILC EB power spectrum and the 2$\sigma$ uncertainties of estimated $\beta$. The optimal uncertainty of the EB power spectrum in the fourth row is computed from Equation~\ref{eq:sigma-dl}.}
    \renewcommand\arraystretch{1.5}
    \begin{tabular}{lccccc}
        \bottomrule
        Case & Baseline & A & C & D & E \\
        \hline
        $f_{\rm sky}$ & 10\% & 10\% & 10\% & 10\% & 5\% \\
        \hline
        Band-combined $\sigma_n^P$ ($\mu$K-arcmin) & 14 & 19 & 10 & 8 & 10 \\
        \hline
        Integration time (module*year)  & 4 & 2 & 8 & 12 & 4 \\
        \hline
        Optimal $\sigma(\hat{\mathcal{D}}_{\ell=107}^{EB,\rm NILC})$ ($\times10^{-3}\mu K^2$) & 6.9 & 9.1 & 5.4 & 4.7 & 7.6 \\
        \hline
        Actual $\sigma(\hat{\mathcal{D}}_{\ell=107}^{EB,\rm NILC})$ ($\times10^{-3}\mu K^2$) & 7.3 & 9.1 & 5.6 & 5.1 & 9.1 \\
        \hline
        TB-EB combined 95\% C.L. of $\beta$ (deg) & $\pm$0.058 & $\pm$0.071 & $\pm$0.046 & $\pm$0.041 & $\pm$0.070 \\
        \toprule
    \end{tabular}
    \label{tab:config}
\end{table}

\begin{figure}[tbp]
    \centering
    \begin{subfigure}[b]{0.6\textwidth}
        \centering
        \includegraphics[width=\textwidth]{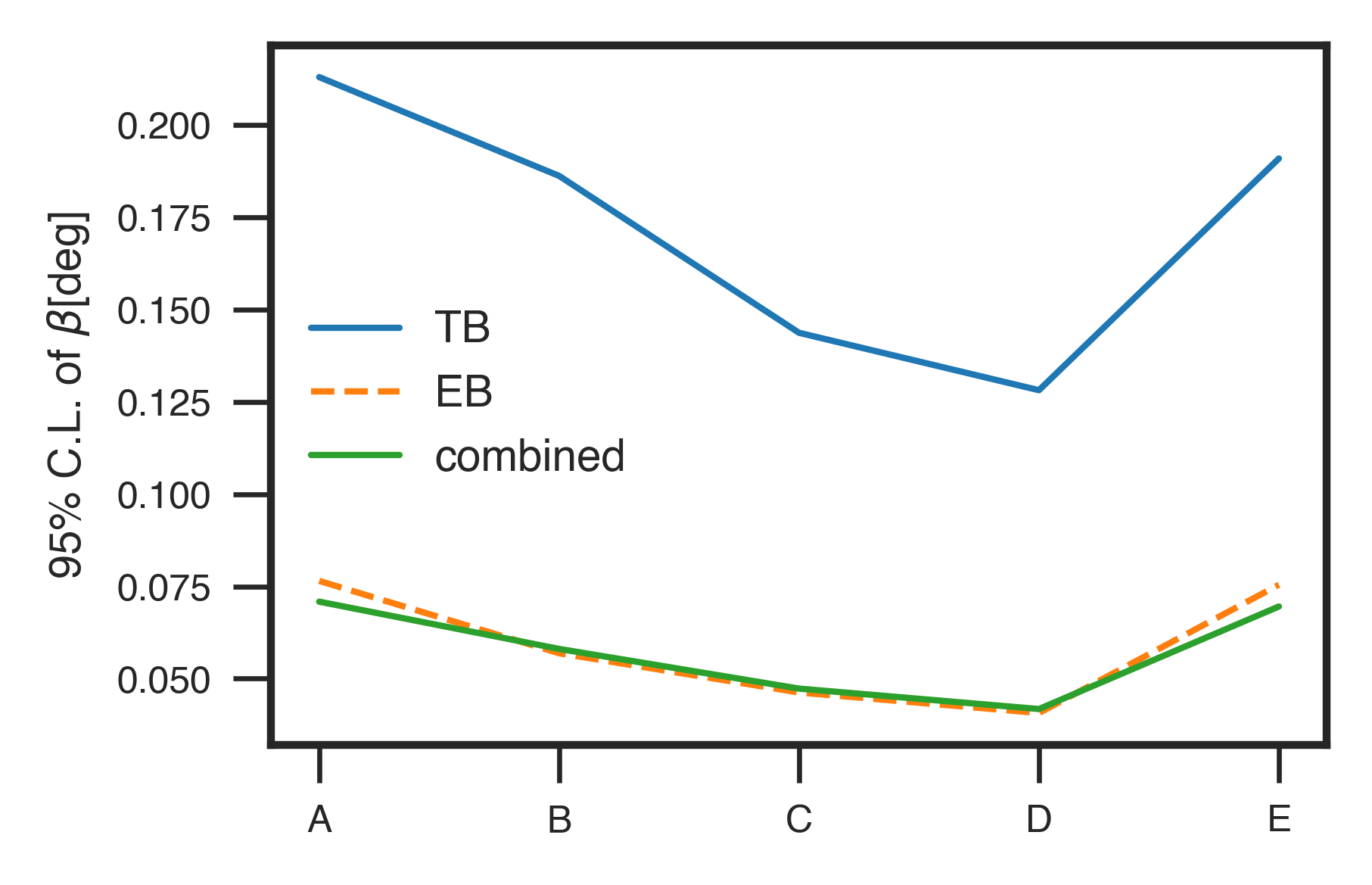}
    \end{subfigure}
    \caption{The 95\% C.L. of the estimated $\beta$ from TB, EB and TB-EB combined power spectra for the five cases listed in Table~\ref{tab:config}. Case `B' denotes the baseline.}
    \label{fig:95CL}
\end{figure}

We apply the same NILC pipeline on simulations of the four new configurations and compute their TB/EB power spectra. To validate the results of the uncertainties of NILC power spectra, we compute the theoretical optimal uncertainties of the TB and EB power spectra without intrinsic cross correlations as:
\begin{equation}\label{eq:sigma-dl}
\begin{aligned}
    \sigma(\hat{\mathcal{D}}_\ell^{TB,\rm NILC})&=\sqrt{\frac{1}{(2\ell+1)\Delta\ell f_{\rm sky}}}
    \sqrt{(\mathcal{D}_\ell^{TT}+\mathcal{N}_\ell^{TT}/b_\ell^{2})(\mathcal{D}_\ell^{BB}+\mathcal{N}_\ell^{BB}/b_\ell^{2})}\,,
    \\
    \sigma(\hat{\mathcal{D}}_\ell^{EB,\rm NILC})&=\sqrt{\frac{1}{(2\ell+1)\Delta\ell f_{\rm sky}}}
    \sqrt{(\mathcal{D}_\ell^{EE}+\mathcal{N}_\ell^{EE}/b_\ell^{2})(\mathcal{D}_\ell^{BB}+\mathcal{N}_\ell^{BB}/b_\ell^{2})}\,,
\end{aligned}
\end{equation}
where $\Delta\ell$ refers to the multipole bin size, $f_{\rm sky}$ refers to the sky coverage, $\mathcal{D}_\ell^{XX},\, X\in\{T,E,B\}$ represents the theoretical CMB power spectrum and $\mathcal{N}_\ell^{XX}/b_\ell^{2}$ represents the noise power spectrum corrected for the Gaussian beam. We compare the optimal uncertainties of the EB power spectrum and the uncertainties from 100 NILC cleaned simulations in Table~\ref{tab:config}. The actual uncertainties are slightly higher than the optimal ones due to the existence of foregrounds and imperfect localization of NILC. Note that the baseline case has a lower uncertainty of EB power than Case `E' though they have the same integration time, since the former has utilized the ancillary data of Planck and WMAP on a larger sky patch than the latter.

We repeat the parameterization analysis in section~\ref{sec:param} on the four new configuration schemes. The results of the 95\% C.L. of $\beta$ are shown in Figure~\ref{fig:95CL} and Table~\ref{tab:config}. Longer integration time will significantly improve the sensitivity of birefringence estimation. With double the survey time, we obtain the 2$\sigma$ uncertainty of $\beta$ as 0.046$^\circ$, and for triple the survey time, it reaches 0.041$^\circ$. Scanning on a smaller sky patch with a fixed survey time would not increase our sensitivity.

\section{Conclusions}
\label{sec:conclusions}
This paper presents the results of TB and EB correlations \revise{using a combination of AliCPT-1, Planck-HFI and WMAP-K} simulations. We use the NILC method to clean the foregrounds, ensuring that the correlations by chance from foregrounds would not raise biases comparable to the uncertainties. To parameterize the uncertainty, we measure the 95\% C.L. threshold of the birefringence angle $|\beta_*|$, the data beyond which could reliably favor the parity violation with $2\sigma$ significance under our sensitivity.

We also compare the performance of NILC and cILC to solely clean the B modes. The cILC is designed to further reduce the residual foregrounds by nulling the dominant foreground components, i.e. thermal dust and synchrotron. In the AliCPT-1 B-mode forecast \cite{ghoshPerformanceForecastsPrimordial2022}, cILC is adopted instead of NILC since NILC leaves a significant bias due to the high foreground residuals. However, for the TB and EB cross spectra, the non-zero correlations from foregrounds are negligible compared to the uncertainties. In such case, NILC wins the competition with its lower uncertainties of the estimated power spectra.

We use a $\chi^2$ test to parameterize the sensitivity of our estimation on TB and EB spectra, by assuming that the possible parity violation is induced by cosmic birefringence. We find that combining TB and EB correlations, the CMB cross power spectrum at $\ell=[40,1000]$ with a birefringence angle $|\beta|>0.058^\circ$ provides an over $2\sigma$ significance on parity violation, after the first AliCPT observing season. There is a caveat that this is not a forecast for estimating the birefringence angle which is commonly estimated from the cross frequency spectra, and the possible systematic errors are not included in our analysis, but it sets a benchmark for the effects of systematics that might significantly bias the estimation. \revise{We further investigate the sensitivity for different survey durations and an alternative scanning strategy.}

The foreground model considered in this work does not take into account the intrinsic TB correlation from thermal dust emissions because the correlation is not robustly estimated till now. The miscalibration angle is degenerate with the birefringence angle in CMB, but Galactic foreground polarization is almost only rotated by the miscalibration angle; thus we could simultaneously determine both of them with the prior of CMB EE and BB power spectra \cite{minamiSimultaneousDeterminationCosmic2019,minamiSimultaneousDeterminationCosmic2020}.
A complete forecast of estimating the birefringence angle needs to be carried out in the future work, with the miscalibration angle and the dust TB correlations considered.

\appendix
\section{Introduction of ILC methods}
\label{sec:methods}
\subsection{Standard ILC}

The Internal Linear Combination (ILC) method \cite{eriksenForegroundRemovalWilkinson2004,tegmarkHighResolutionForeground2003} is a blind component separation technique widely used in recovering the CMB signal from the multi-frequency sky observations without any assumption of foreground emissions and noise. The ILC cleaned CMB map is constructed as a linear combination of the multi-frequency data through minimizing its variance with remaining the CMB component unchanged.

First, the observed sky of each frequency band can be modeled as the sum of the CMB anisotropies $s_{\rm CMB}$, foregrounds $f_\nu$ and instrumental noise $n_\nu$ (each is a map in either pixel space or harmonic space):
\begin{equation}
    d_\nu = a_\nu s_{\rm CMB} + f_\nu + n_\nu\,,
\end{equation}
where $a_\nu$ is the spectral response of CMB at frequency $\nu$. Here we assume that the CMB signal is frequency independent after calibration with respect to CMB, i.e. $a_\nu=1$. For simplicity, we recast the maps at $n_\nu$ frequency channels in a $n_\nu\times1$ column vector form:
\begin{equation}
    \boldsymbol{d} = \boldsymbol{a}s_{\rm CMB} + \boldsymbol{f} + \boldsymbol{n}\,,
    \label{eq:model}
\end{equation}
where $\boldsymbol{d}=[d_\nu]^T$ is the $n_\nu\times1$ column vector of the observed data, and $\boldsymbol{a}=[a_\nu]^T$ is the CMB spectral energy distribution (SED) vector. Hereafter, the observed data is already re-convolved to a common beam and resolution: $d_{\ell m}^\nu=\frac{b_\ell^0}{b_\ell^\nu}d_{\ell m}^{\nu,obs}$, where $b_\ell^{0}$ is the common beam transfer function and $b_\ell^{\nu}$ is the beam function for each frequency band.  $d_{\ell m}^\nu$ is the harmonic coefficients of $d_\nu$, and $d_{\ell m}^{\nu,obs}$ is the harmonic coefficients of the raw data.

The estimated CMB map is defined as a linear combination of all observed frequency maps with a constraint of the frequency weights $\boldsymbol{w}^T \boldsymbol{a}=1$:
\begin{equation}
    \hat s_{\rm CMB} = \boldsymbol{w}^T \boldsymbol{d}
    = s_{\rm CMB} + \boldsymbol{w}^T \boldsymbol{f} + \boldsymbol{w}^T \boldsymbol{n}\,,
    \label{eq:ilc-cleaned-map}
\end{equation}
where $\boldsymbol{w}=[w_\nu]^T$ is a $n_\nu\times1$ vector of weights. The weights $\boldsymbol{w}$ are computed for each pixel (or harmonic multipole) $p$ by minimizing the ensemble variance of the ILC-cleaned map $\langle\hat s^2_{\rm CMB} (p)\rangle$ using the Lagrange multiplier method, so that the variance of the error $(\hat s_{\rm CMB}-s_{\rm CMB})$ is also minimized. Since we have only one observation of the universe, in pixel domain, the variance localized at pixel $p$ is estimated empirically by the variance of the observed data over the vicinity of the pixel $p$. In harmonic space, the variance at angular scale $\ell$ is estimated by the angular power spectrum averaged over a multipole bin centered at $\ell$. The ILC weights in pixel space are given as:
\begin{equation}
    \boldsymbol{w}(p) = \frac{\boldsymbol {\hat C}(p)^{-1} \boldsymbol a}{\boldsymbol a^T\boldsymbol {\hat C}(p)^{-1} \boldsymbol a}\, ,\ \text{where}\ \boldsymbol{\hat C}(p)=\frac{1}{N_{p'}}\sum_{p'\in V(p)}q(p',p)\boldsymbol{d}(p')^\dagger \boldsymbol{d}(p')\,,
    \label{eq:pilc-wgts}
\end{equation}
where the $n_\nu\times n_\nu$ data covariance matrix $\boldsymbol {\hat C}$ is an estimate of the real ensemble covariance matrix, $V(p)$ is the vicinity of the pixel $p$, and $q(p',p)$ is some kind of weight with respective to neighbouring pixels $p'$ and $p$ (for instance a Gaussian smooth kernel centered at pixel $p$).

Likewise, in harmonic space, the ILC weights are given by:
\begin{equation}
    \boldsymbol{w}_{\ell} = \frac{\boldsymbol {\hat C}_\ell^{-1} \boldsymbol a}{\boldsymbol a^T\boldsymbol {\hat C}_\ell^{-1} \boldsymbol a}\, ,\ \text{where}\ \boldsymbol{\hat C}_\ell=\sum_{\ell'\in B(\ell)}\sum_{m=-\ell'}^{\ell'}\boldsymbol{d}_{\ell'm}^\dagger \boldsymbol{d}_{\ell'm}\,,
    \label{eq:hilc-wgts}
\end{equation}
where the covariance matrix $\boldsymbol{\hat C}_\ell$ is summed over the multipole bin $B(\ell)$ centered at $\ell$. The final foreground-cleaned CMB map is $\hat s_{\ell m}=\boldsymbol{w}^T_{\ell}\boldsymbol{d}_{\ell m}$.

\subsection{NILC}
\label{sec:nilc}
The needlet ILC (NILC) has been used to extract CMB anisotropies for numerous CMB experiments since WMAP \cite{basakNeedletILCAnalysis2012,basakNeedletILCAnalysis2013,planckcollaborationPlanck2018Results2020a,remazeillesExploringCosmicOrigins2017,remazeillesPeelingForegroundsConstrained2021,adakBModeForecastCMBBh2021,krachmalnicoffInflightPolarizationAngle2022,zhangEfficientILCAnalysis2022,caronesAnalysisNILCPerformance2023, aurlienForegroundSeparationConstraints2023,fuskelandTensortoscalarRatioForecasts2023}.
In NILC, each frequency map is filtered by a set of needlet bands in harmonic domain, and the ILC procedure is implemented independently on each pixel for each filtered map. The observed data filtered by the needlet bands, i.e the needlet maps are:
\begin{equation}
    d^{\nu, j}_{\ell m} = h^j_\ell d^\nu_{\ell m}\,,
\end{equation}
where the needlet window functions $h^j_\ell$ satisfying $\sum_j (h^j_\ell)^2=1$ take charge of the localization in harmonic space. In this work we adopt the cosine window functions which can be written as:
\begin{equation}
    h_\ell^j=\left\{
        \begin{aligned}
           &\cos(\frac{\pi}{2}\frac{\ell_{mid}^j-\ell}{\ell_{mid}^j-\ell_{mid}^{j-1}}) \,, \ell_{mid}^{j-1}\leq\ell<\ell_{mid}^j \,,\\
           &\cos(\frac{\pi}{2}\frac{\ell-\ell_{mid}^j}{\ell_{mid}^{j+1}-\ell_{mid}^j}) \,, \ell_{mid}^j\leq\ell<\ell_{mid}^{j+1} \,,\\
           & 0 \,, \text{otherwise}\,,
        \end{aligned}
    \right.
\end{equation}
where $j\in\{1,2,\dots,n_j\}$, $\ell_{mid}^j$ is the center multipole of the $j$-th needlet band, $h_\ell^1=1$ when $0\leq\ell<\ell_{mid}^1$, and $h_\ell^{n_j}=1$ when $\ell\geq\ell_{mid}^{n_j}$. The needlet maps are then transformed into pixel domain: 
\begin{equation}
    b^\nu_j(n_{jk}) = \sum_{\ell m} h^j_\ell d^\nu_{\ell m} Y_{\ell m}(n_{jk})\,.
\end{equation}
where $n_{jk}$ represents the $k$-th pixel of the $j$-th needlet map.

In order to estimate the covariance matrices for the $k$-th pixel of the $j$-th needlet scale $\boldsymbol C_{jk} = C^{\nu_1\times \nu_2}_{jk} = \langle b^{\nu_1}_j(n_{jk}) b^{\nu_2}_j(n_{jk}) \rangle$, we compute the empirical covariance $\boldsymbol{\hat C}_{jk}$ by averaging the needlet coefficient products $b^\nu_j(n_{jk}) b^\nu_j(n_{jk})$ over a sky patch around the $k$-th pixel, which can be written as:
\begin{equation}\label{eq:nilc-cov}
    \hat C^{\nu_1\times \nu_2}_{jk} = \frac{1}{N_k} \sum_{k'} q_j(k, k') b^{\nu_1}_j(n_{jk'}) b^{\nu_2}_j(n_{jk'})\,,
\end{equation}
where $q_j(k, k')$ are weights dependent on the needlet scale, and $N_k$ is the number of pixels averaged out. As the standard ILC method, the NILC weights are given by:
\begin{equation}
    \boldsymbol{w}^{\rm NILC}_{j}(n_{jk}) = \frac{\boldsymbol {\hat C}_{jk}^{-1} \boldsymbol a}{\boldsymbol a^T\boldsymbol {\hat C}_{jk}^{-1} \boldsymbol a}\,.
    \label{eq:nilc-wgts}
\end{equation}

Each needlet map after foreground cleaning is the linear combination of the input multi-frequency needlet maps for each needlet band $j$. The NILC cleaned map is transformed from the needlet space into pixel space:
\begin{subequations}
    \begin{align}
    &b^{\rm NILC}_j(n_{jk}) = \sum_{\nu} w^{\rm NILC}_{\nu, j}(n_{jk}) b^\nu_j(n_{jk})\,,
    \\
    &\hat s^{\rm NILC}_{\ell m} = \sum_{jk}b^{\rm NILC}_j(n_{jk}) \frac{4\pi}{N_{\rm{pix}}^j} h^j_\ell Y^*_{\ell m}(n_{jk})\,,
    \\
    &\hat s_{\rm NILC} (p)= \sum_{\ell m} \hat s^{\rm NILC}_{\ell m} Y_{\ell m} (p)\,.
    \end{align}
\end{subequations}

\subsection{Constrained ILC}
\label{sec:cilc}
The constrained ILC (cILC) method \cite{remazeillesCMBSZEffect2011,remazeillesPeelingForegroundsConstrained2021} further introduces assumptions of the spectral energy density (SED) shapes of the dominated foreground components, i.e. Galactic thermal dust and synchrotron emissions, for CMB polarization analysis. It imposes extra constraints on the weights to cancel those unwanted components. Compared to the standard ILC, this method significantly reduces the residual foregrounds but increases the residual noise level as an expense.

Given the modeled SEDs of foreground components, the observations are written as:
\begin{equation}
    \begin{aligned}
    \boldsymbol{d} = & \boldsymbol{a}_{\rm CMB}s_{\rm CMB} + \sum_{i=1}^2\boldsymbol{a}_i s_i + \boldsymbol{n}\\
    = & \boldsymbol{A}\boldsymbol{s} + \boldsymbol{n}\,,
    \end{aligned}
\end{equation}
where $s_1$ and $s_2$ are thermal dust and synchrotron templates whose SED vectors are $\boldsymbol{a}_1$ and $\boldsymbol{a}_2$, respectively. The $n_\nu\times3$ matrix $\boldsymbol{A}=\begin{bmatrix}\boldsymbol{a}_{\rm CMB} & \boldsymbol{a}_1 & \boldsymbol{a}_2\end{bmatrix}$ is the so-called mixing matrix and the $3\times1$ vector $\boldsymbol{s}=\begin{bmatrix}s_{\rm CMB} & s_1 & s_2\end{bmatrix}^T$ collects the templates of three components.

To eliminate the foreground components with the given SEDs, the constraints are set as:
\begin{equation}
    \boldsymbol{w}^T \boldsymbol{A} = \boldsymbol{e}^T\,,
\end{equation}
where $\boldsymbol{e}=\begin{bmatrix}1 & 0 & 0\end{bmatrix}$. The cILC weights are then given by:
\begin{equation}
    \boldsymbol{w}^{T} = \boldsymbol{e}^T \left( \boldsymbol{A}^T\hat{\boldsymbol{C}}^{-1} \boldsymbol{A}\right)^{-1}\boldsymbol{A}^T \hat{\boldsymbol{C}}^{-1}\,,
\end{equation}
where $\hat{\boldsymbol{C}}$ is the data covariance matrix. The cILC cleaned CMB map is written as $\hat s_{\rm CMB}=\boldsymbol{w}^T\cdot \boldsymbol{d}$. In this work, we use the cILC implemented in harmonic domain since we obtain similar results in needlet domain.

The mixing matrix $\boldsymbol{A}$ is obtained from the assumed SEDs of the synchrotron and thermal dust components. We adopt the SED parameters from the Planck constraints on polarized foregrounds \cite{planckcollaborationPlanck2018Results2020e,planckcollaborationPlanckIntermediateResults2016a}. The SED of CMB anisotropies is modeled as a differential black body, of synchrotron emission as a power law with a fixed spectral index of $\beta_{\rm sync}=-3$, and of the thermal dust emission as a modified black body with the dust temperature of $T_{\rm dust}=19.6$ K and the dust spectral index of $\beta_{\rm dust}=1.59$, for data in Rayleigh-Jeans brightness temperature units. The SEDs after calibration to the CMB thermodynamic temperature are written as:
\begin{equation}
    \left\{
        \begin{aligned}
            & a_{\rm CMB}(\nu)=1\,,\\
            & a_{\rm sync}(\nu)=\frac{g_\nu}{g_{\nu_s}}(\frac{\nu}{\nu_s})^{\beta_s} \,,\\
            & a_{\rm dust}(\nu)=\frac{g_\nu}{g_{\nu_d}}(\frac{\nu}{\nu_d})^{\beta_d+1}\frac{\exp{(x_d(\nu_d))}-1}{\exp{(x_d(\nu))}-1}\,,
        \end{aligned}
    \right.
\end{equation}
where $x_d(\nu)\equiv\frac{h\nu}{k_B T_d}$, $g_\nu$ is the unit conversion factor from $\mu K_{\rm RJ}$ to $\mu K_{\rm CMB}$, $\nu_s=23$ GHz and $\nu_d=353$ GHz are the reference frequencies.

\section{\revise{Summary of the template cleaning method}}
\label{sec:tc}
The template cleaning method \cite{liuMethodsPixelDomain2019} is a pixel-based technique to remove the E-B leakage due to the E-B mixing in partial-sky polarization decomposition \cite{zhaoSeparatingTypesPolarization2010,ghoshEndingPartialSky2021}. With the masked QU maps, we perform the template cleaning as following steps:

\begin{enumerate}
    \item The partial-sky polarization fields $\boldsymbol{P}=(Q,U)$ can be decomposed into the so-called E and B families: $\boldsymbol{P_E}=(Q_E,U_E)$ and $\boldsymbol{P_B}=(Q_B,U_B)$, which is realized by $(Q,U)\rightarrow E;B\rightarrow (Q_E,U_E);(Q_B,U_B)$. Here we directly decompose the polarization fields into E and B modes (though E-B mixed), and transform the E and B modes into polarization fields respectively.
    \item Likewise, we decompose the masked E family $\boldsymbol{P_E}$ into its E and B families $\boldsymbol{P'_E},\,\boldsymbol{P'_B}$.
    \item We take the masked $\boldsymbol{P'_B}$ as an E-B leakage template to linearly fit the leakage in the masked $\boldsymbol{P_B}$. Finally we subtract the fitted leakage from the masked $\boldsymbol{P_B}$ to obtain the pure B-mode map, i.e. $\boldsymbol{P_B}^{\rm pure}=\boldsymbol{P_B}-\alpha\boldsymbol{P'_B}$ where $\alpha$ is the linear coefficient.
\end{enumerate}

\acknowledgments
We appreciate the helpful discussions in the AliCPT science
team.
This work is supported by the National Key R\&D Program of China Grant No. 2021YFC2203102 and 2022YFC2204602, Strategic Priority Research Program of the Chinese
Academy of Science Grant No. XDB0550300,
NSFC No. 12325301 and 12273035, the China Manned Space Project with No.CMS-CSST-2021-B01, the China Manned Space Program through its Space Application System,
and the 111 Project for "Observational and Theoretical Research on Dark Matter and Dark Energy" (B23042). L.S. is supported by the National Key R\&D Program of China (2020YFC2201600) and NSFC grant 12150610459.


    
    


\bibliography{Bib, Bib2}

\providecommand{\href}[2]{#2}\begingroup\raggedright\begin{thebibliography}{10}

\bibitem{zaldarriagaAllSkyAnalysisPolarization1997}
M.~Zaldarriaga and U.~Seljak, \emph{An {{All-Sky Analysis}} of {{Polarization}}
  in the {{Microwave Background}}},
  \href{https://doi.org/10.1103/PhysRevD.55.1830}{\emph{Phys. Rev. D}
  {\bfseries 55} (1997) 1830}
  [\href{https://arxiv.org/abs/astro-ph/9609170}{{\ttfamily
  astro-ph/9609170}}].

\bibitem{seljakSignatureGravityWaves1997}
U.~Seljak and M.~Zaldarriaga, \emph{Signature of {{Gravity Waves}} in
  {{Polarization}} of the {{Microwave Background}}},
  \href{https://doi.org/10.1103/PhysRevLett.78.2054}{\emph{Phys. Rev. Lett.}
  {\bfseries 78} (1997) 2054}
  [\href{https://arxiv.org/abs/astro-ph/9609169}{{\ttfamily
  astro-ph/9609169}}].

\bibitem{lueCosmologicalSignatureNew1999}
A.~Lue, L.~Wang and M.~Kamionkowski, \emph{Cosmological {{Signature}} of {{New
  Parity-Violating Interactions}}},
  \href{https://doi.org/10.1103/PhysRevLett.83.1506}{\emph{Phys. Rev. Lett.}
  {\bfseries 83} (1999) 1506}
  [\href{https://arxiv.org/abs/astro-ph/9812088}{{\ttfamily
  astro-ph/9812088}}].

\bibitem{fengSearchingViolationCosmic2006}
B.~Feng, M.~Li, J.-Q.~Xia, X.~Chen and X.~Zhang, \emph{Searching for {{C P T
  Violation}} with {{Cosmic Microwave Background Data}} from {{WMAP}} and
  {{BOOMERANG}}},
  \href{https://doi.org/10.1103/PhysRevLett.96.221302}{\emph{Phys. Rev. Lett.}
  {\bfseries 96} (2006) 221302}.

\bibitem{quadcollaborationParityViolationConstraints2009}
Q.~Collaboration, E.Y.S.~Wu, P.~Ade, J.~Bock, M.~Bowden, M.L.~Brown et~al.,
  \emph{Parity violation constraints using 2006-2007 {{QUaD CMB}} polarization
  spectra}, \href{https://doi.org/10.1103/PhysRevLett.102.161302}{\emph{Phys.
  Rev. Lett.} {\bfseries 102} (2009) 161302}
  [\href{https://arxiv.org/abs/0811.0618}{{\ttfamily 0811.0618}}].

\bibitem{leeQuestionParityConservation1956}
T.D.~Lee and C.N.~Yang, \emph{Question of {{Parity Conservation}} in {{Weak
  Interactions}}}, \href{https://doi.org/10.1103/PhysRev.104.254}{\emph{Phys.
  Rev.} {\bfseries 104} (1956) 254}.

\bibitem{wuExperimentalTestParity1957}
C.S.~Wu, E.~Ambler, R.W.~Hayward, D.D.~Hoppes and R.P.~Hudson,
  \emph{Experimental {{Test}} of {{Parity Conservation}} in {{Beta Decay}}},
  \href{https://doi.org/10.1103/PhysRev.105.1413}{\emph{Phys. Rev.} {\bfseries
  105} (1957) 1413}.

\bibitem{alexanderChernSimonsModifiedGeneral2009}
S.~Alexander and N.~Yunes, \emph{Chern-{{Simons Modified General Relativity}}},
  \href{https://doi.org/10.1016/j.physrep.2009.07.002}{\emph{Physics Reports}
  {\bfseries 480} (2009) 1} [\href{https://arxiv.org/abs/0907.2562}{{\ttfamily
  0907.2562}}].

\bibitem{marshAxionCosmology2016}
D.J.E.~Marsh, \emph{Axion cosmology},
  \href{https://doi.org/10.1016/j.physrep.2016.06.005}{\emph{Phys. Rept.}
  {\bfseries 643} (2016) 1} [\href{https://arxiv.org/abs/1510.07633}{{\ttfamily
  1510.07633}}].

\bibitem{ferreiraUltraLightDarkMatter2021}
E.G.M.~Ferreira, \emph{Ultra-{{Light Dark Matter}}},
  \href{https://doi.org/10.1007/s00159-021-00135-6}{\emph{Astron. Astrophys.
  Rev.} {\bfseries 29} (2021) 7}
  [\href{https://arxiv.org/abs/2005.03254}{{\ttfamily 2005.03254}}].

\bibitem{yadavProbingPrimordialMagnetism2012}
A.~Yadav, L.~Pogosian and T.~Vachaspati, \emph{Probing primordial magnetism
  with off-diagonal correlators of {{CMB}} polarization},
  \href{https://doi.org/10.1103/PhysRevD.86.123009}{\emph{Phys. Rev. D}
  {\bfseries 86} (2012) 123009}
  [\href{https://arxiv.org/abs/1207.3356}{{\ttfamily 1207.3356}}].

\bibitem{deCMBFaradayRotation2013}
S.~De, L.~Pogosian and T.~Vachaspati, \emph{{{CMB Faraday}} rotation as seen
  through the {{Milky Way}}},
  \href{https://doi.org/10.1103/PhysRevD.88.063527}{\emph{Phys. Rev. D}
  {\bfseries 88} (2013) 063527}
  [\href{https://arxiv.org/abs/1305.7225}{{\ttfamily 1305.7225}}].

\bibitem{carrollLimitsLorentzParityviolating1990}
S.M.~Carroll, G.B.~Field and R.~Jackiw, \emph{Limits on a {{Lorentz-}} and
  parity-violating modification of electrodynamics},
  \href{https://doi.org/10.1103/PhysRevD.41.1231}{\emph{Phys. Rev. D}
  {\bfseries 41} (1990) 1231}.

\bibitem{zhaoFluctuationsCosmologicalBirefringence2014}
W.~Zhao and M.~Li, \emph{Fluctuations of cosmological birefringence and the
  effect on {{CMB B-mode}} polarization},
  \href{https://doi.org/10.1103/PhysRevD.89.103518}{\emph{Phys. Rev. D}
  {\bfseries 89} (2014) 103518}
  [\href{https://arxiv.org/abs/1403.3997}{{\ttfamily 1403.3997}}].

\bibitem{zhaoDetectingRelicGravitational2014}
W.~Zhao and M.~Li, \emph{Detecting relic gravitational waves in the {{CMB}}:
  {{The}} contamination caused by the cosmological birefringence},
  \href{https://doi.org/10.1016/j.physletb.2014.09.003}{\emph{Physics Letters
  B} {\bfseries 737} (2014) 329}
  [\href{https://arxiv.org/abs/1402.4324}{{\ttfamily 1402.4324}}].

\bibitem{eskiltFrequencyDependentConstraintsCosmic2022}
J.R.~Eskilt, \emph{Frequency-{{Dependent Constraints}} on {{Cosmic
  Birefringence}} from the {{LFI}} and {{HFI Planck Data Release}} 4},
  \href{https://doi.org/10.1051/0004-6361/202243269}{\emph{A\&A} {\bfseries
  662} (2022) A10} [\href{https://arxiv.org/abs/2201.13347}{{\ttfamily
  2201.13347}}].

\bibitem{eskiltImprovedConstraintsCosmic2022}
J.R.~Eskilt and E.~Komatsu, \emph{Improved {{Constraints}} on {{Cosmic
  Birefringence}} from the {{WMAP}} and {{Planck Cosmic Microwave Background
  Polarization Data}}},
  \href{https://doi.org/10.1103/PhysRevD.106.063503}{\emph{Phys. Rev. D}
  {\bfseries 106} (2022) 063503}
  [\href{https://arxiv.org/abs/2205.13962}{{\ttfamily 2205.13962}}].

\bibitem{pgw2}
A.~{Wang}, Q.~{Wu}, W.~{Zhao} and T.~{Zhu}, \emph{{Polarizing primordial
  gravitational waves by parity violation}},
  \href{https://doi.org/10.1103/PhysRevD.87.103512}{\emph{Phys. Rev. D}
  {\bfseries 87} (2013) 103512}
  [\href{https://arxiv.org/abs/1208.5490}{{\ttfamily 1208.5490}}].

\bibitem{pgw3}
J.~{Qiao}, T.~{Zhu}, W.~{Zhao} and A.~{Wang}, \emph{{Polarized primordial
  gravitational waves in the ghost-free parity-violating gravity}},
  \href{https://doi.org/10.1103/PhysRevD.101.043528}{\emph{Phys. Rev. D}
  {\bfseries 101} (2020) 043528}
  [\href{https://arxiv.org/abs/1911.01580}{{\ttfamily 1911.01580}}].

\bibitem{pgw4}
J.~{Qiao}, Z.~{Li}, T.~{Zhu}, R.~{Ji}, G.~{Li} and W.~{Zhao}, \emph{{Testing
  parity symmetry of gravity with gravitational waves}},
  \href{https://doi.org/10.3389/fspas.2022.1109086}{\emph{Frontiers in
  Astronomy and Space Sciences} {\bfseries 9} (2023) 442}
  [\href{https://arxiv.org/abs/2211.16825}{{\ttfamily 2211.16825}}].

\bibitem{pgw5}
T.-C.~{Li}, T.~{Zhu}, W.~{Zhao} and A.~{Wang}, \emph{{Power spectra and
  circular polarization of primordial gravitational waves with parity and
  Lorentz violations}},
  \href{https://doi.org/10.48550/arXiv.2403.05841}{\emph{arXiv e-prints} (2024)
  arXiv:2403.05841} [\href{https://arxiv.org/abs/2403.05841}{{\ttfamily
  2403.05841}}].

\bibitem{collaborationBICEPKeckXVII2023}
{\scshape BICEP/Keck} collaboration, \emph{{{BICEP}} / {{Keck XVII}}: {{Line}}
  of {{Sight Distortion Analysis}}: {{Estimates}} of {{Gravitational Lensing}},
  {{Anisotropic Cosmic Birefringence}}, {{Patchy Reionization}}, and
  {{Systematic Errors}}},
  \href{https://doi.org/10.3847/1538-4357/acc85c}{\emph{Astrophys. J.}
  {\bfseries 949} (2023) 43}
  [\href{https://arxiv.org/abs/2210.08038}{{\ttfamily 2210.08038}}].

\bibitem{sheehyDeprojectingBeamSystematics2019}
C.~Sheehy, \emph{Deprojecting beam systematics for next-generation {{CMB
  B-mode}} searches},
  \href{https://doi.org/10.48550/arXiv.1911.03547}{\emph{arXiv:1911.03547
  [astro-ph.IM]} (2019) } [\href{https://arxiv.org/abs/1911.03547}{{\ttfamily
  1911.03547}}].

\bibitem{johnsonMAXIPOLCosmicMicrowave2007}
B.R.~Johnson, J.~Collins, M.E.~Abroe, P.A.R.~Ade, J.~Bock, J.~Borrill et~al.,
  \emph{{{MAXIPOL}}: {{Cosmic Microwave Background Polarimetry Using}} a
  {{Rotating Half-Wave Plate}}},
  \href{https://doi.org/10.1086/518105}{\emph{ApJ} {\bfseries 665} (2007) 42}
  [\href{https://arxiv.org/abs/astro-ph/0611394}{{\ttfamily
  astro-ph/0611394}}].

\bibitem{patanchonEffectInstrumentalPolarization2023}
G.~Patanchon, H.~Imada, H.~Ishino and T.~Matsumura, \emph{Effect of
  {{Instrumental Polarization}} with a {{Half-Wave Plate}} on the
  ${{B}}$-{{Mode Signal}}: {{Prediction}} and {{Correction}}},
  \href{https://doi.org/10.48550/arXiv.2308.00967}{\emph{arXiv:2308.00967
  [astro-ph.CO]} (2023) } [\href{https://arxiv.org/abs/2308.00967}{{\ttfamily
  2308.00967}}].

\bibitem{monelliImpactHalfwavePlate2022}
M.~Monelli, E.~Komatsu, A.E.~Adler, M.~Billi, P.~Campeti, N.~Dachlythra et~al.,
  \emph{Impact of half-wave plate systematics on the measurement of cosmic
  birefringence from {{CMB}} polarization},
  \href{https://arxiv.org/abs/2211.05685}{{\ttfamily 2211.05685}}.

\bibitem{monelliImpactHalfwavePlate2023}
M.~Monelli, E.~Komatsu, T.~Ghigna, T.~Matsumura, G.~Pisano and R.~Takaku,
  \emph{Impact of half-wave plate systematics on the measurement of {{CMB}}
  ${{B}}$-mode polarization},
  \href{https://arxiv.org/abs/2311.07999}{{\ttfamily 2311.07999}}.

\bibitem{minamiSimultaneousDeterminationCosmic2019}
Y.~Minami, H.~Ochi, K.~Ichiki, N.~Katayama, E.~Komatsu and T.~Matsumura,
  \emph{Simultaneous determination of the cosmic birefringence and
  miscalibrated polarisation angles from {{CMB}} experiments},
  \href{https://doi.org/10.1093/ptep/ptz079}{\emph{PTEP} {\bfseries 2019}
  (2019) 083E02} [\href{https://arxiv.org/abs/1904.12440}{{\ttfamily
  1904.12440}}].

\bibitem{minamiSimultaneousDeterminationCosmic2020}
Y.~Minami and E.~Komatsu, \emph{Simultaneous determination of the cosmic
  birefringence and miscalibrated polarization angles {{II}}: {{Including}}
  cross frequency spectra},
  \href{https://doi.org/10.1093/ptep/ptaa130}{\emph{PTEP} {\bfseries 2020}
  (2020) 103E02} [\href{https://arxiv.org/abs/2006.15982}{{\ttfamily
  2006.15982}}].

\bibitem{planckcollaborationPlanck2018Results2020e}
P.~Collaboration, Y.~Akrami, M.~Ashdown, J.~Aumont, C.~Baccigalupi,
  M.~Ballardini et~al., \emph{Planck 2018 results. {{XI}}. {{Polarized}} dust
  foregrounds},
  \href{https://doi.org/10.1051/0004-6361/201832618}{\emph{Astron. Astrophys.}
  {\bfseries 641} (2020) A11}
  [\href{https://arxiv.org/abs/1801.04945}{{\ttfamily 1801.04945}}].

\bibitem{huffenbergerPowerSpectraPolarized2020}
K.M.~Huffenberger, A.~Rotti and D.C.~Collins, \emph{The {{Power Spectra}} of
  {{Polarized}}, {{Dusty Filaments}}},
  \href{https://doi.org/10.3847/1538-4357/ab9df9}{\emph{ApJ} {\bfseries 899}
  (2020) 31} [\href{https://arxiv.org/abs/1906.10052}{{\ttfamily 1906.10052}}].

\bibitem{clarkOriginParityViolation2021}
S.E.~Clark, C.-G.~Kim, J.C.~Hill and B.S.~Hensley, \emph{The {{Origin}} of
  {{Parity Violation}} in {{Polarized Dust Emission}} and {{Implications}} for
  {{Cosmic Birefringence}}},
  \href{https://doi.org/10.3847/1538-4357/ac0e35}{\emph{Astrophys. J.}
  {\bfseries 919} (2021) 53}
  [\href{https://arxiv.org/abs/2105.00120}{{\ttfamily 2105.00120}}].

\bibitem{liProbingPrimordialGravitational2019}
H.~Li, S.-Y.~Li, Y.~Liu, Y.-P.~Li, Y.~Cai, M.~Li et~al., \emph{Probing
  {{Primordial Gravitational Waves}}: {{Ali CMB Polarization Telescope}}},
  \href{https://doi.org/10.1093/nsr/nwy019}{\emph{Natl. Sci. Rev.} {\bfseries
  6} (2019) 145} [\href{https://arxiv.org/abs/1710.03047}{{\ttfamily
  1710.03047}}].

\bibitem{ghoshPerformanceForecastsPrimordial2022}
S.~{Ghosh}, Y.~{Liu}, L.~{Zhang}, S.~{Li}, J.~{Zhang}, J.~{Wang} et~al.,
  \emph{{Performance forecasts for the primordial gravitational wave detection
  pipelines for AliCPT-1}},
  \href{https://doi.org/10.1088/1475-7516/2022/10/063}{\emph{JCAP} {\bfseries
  2022} (2022) 063} [\href{https://arxiv.org/abs/2205.14804}{{\ttfamily
  2205.14804}}].

\bibitem{chen01}
J.~{Chen}, S.~{Ghosh}, H.~{Liu}, L.~{Santos}, W.~{Fang}, S.~{Li} et~al.,
  \emph{{Fast Scalar Quadratic Maximum Likelihood Estimators for the CMB B-mode
  Power Spectrum}}, \href{https://doi.org/10.3847/1538-4365/ac18c9}{\emph{ApJS}
  {\bfseries 257} (2021) 27}
  [\href{https://arxiv.org/abs/2104.07408}{{\ttfamily 2104.07408}}].

\bibitem{delabrouilleFullSkyLow2009}
J.~Delabrouille, J.-F.~Cardoso, M.L.~Jeune, M.~Betoule, G.~Fay and F.~Guilloux,
  \emph{A full sky, low foreground, high resolution {{CMB}} map from {{WMAP}}},
  \href{https://doi.org/10.1051/0004-6361:200810514}{\emph{Astron. Astrophys.}
  {\bfseries 493} (2009) 835}
  [\href{https://arxiv.org/abs/0807.0773}{{\ttfamily 0807.0773}}].

\bibitem{remazeillesCMBSZEffect2011}
M.~Remazeilles, J.~Delabrouille and J.-F.~Cardoso, \emph{{{CMB}} and {{SZ}}
  effect separation with {{Constrained Internal Linear Combinations}}},
  \href{https://doi.org/10.1111/j.1365-2966.2010.17624.x}{\emph{Mon. Not. R.
  Astron. Soc.} {\bfseries 410} (2011) 2481}
  [\href{https://arxiv.org/abs/1006.5599}{{\ttfamily 1006.5599}}].

\bibitem{remazeillesPeelingForegroundsConstrained2021}
M.~Remazeilles, A.~Rotti and J.~Chluba, \emph{Peeling off foregrounds with the
  constrained moment {{ILC}} method to unveil primordial {{CMB B-modes}}},
  \href{https://arxiv.org/abs/2006.08628}{{\ttfamily 2006.08628}}.

\bibitem{gorskiHEALPixFrameworkHigh2005}
K.M.~Gorski, E.~Hivon, A.J.~Banday, B.D.~Wandelt, F.K.~Hansen, M.~Reinecke
  et~al., \emph{{{HEALPix}} -- a {{Framework}} for {{High Resolution
  Discretization}}, and {{Fast Analysis}} of {{Data Distributed}} on the
  {{Sphere}}}, \href{https://doi.org/10.1086/427976}{\emph{ApJ} {\bfseries 622}
  (2005) 759} [\href{https://arxiv.org/abs/astro-ph/0409513}{{\ttfamily
  astro-ph/0409513}}].

\bibitem{planckcollaborationPlanck2018Results2020c}
P.~Collaboration, Y.~Akrami, F.~Arroja, M.~Ashdown, J.~Aumont, C.~Baccigalupi
  et~al., \emph{Planck 2018 results. {{I}}. {{Overview}} and the cosmological
  legacy of {{Planck}}},
  \href{https://doi.org/10.1051/0004-6361/201833880}{\emph{Astron. Astrophys.}
  {\bfseries 641} (2020) A1}
  [\href{https://arxiv.org/abs/1807.06205}{{\ttfamily 1807.06205}}].

\bibitem{lewisEfficientComputationCMB2000}
A.~Lewis, A.~Challinor and A.~Lasenby, \emph{Efficient {{Computation}} of
  {{CMB}} anisotropies in closed {{FRW}} models},
  \href{https://doi.org/10.1086/309179}{\emph{ApJ} {\bfseries 538} (2000) 473}
  [\href{https://arxiv.org/abs/astro-ph/9911177}{{\ttfamily
  astro-ph/9911177}}].

\bibitem{aghanimPlanck2018Results2020a}
{\scshape Planck} collaboration, \emph{Planck 2018 results. {{VI}}.
  {{Cosmological}} parameters},
  \href{https://doi.org/10.1051/0004-6361/201833910}{\emph{Astron. Astrophys.}
  {\bfseries 641} (2020) A6}
  [\href{https://arxiv.org/abs/1807.06209}{{\ttfamily 1807.06209}}].

\bibitem{reineckeImprovedCMBLensing2023}
M.~Reinecke, S.~Belkner and J.~Carron, \emph{Improved {{CMB}} (de-)lensing
  using general spherical harmonic transforms},
  \href{https://doi.org/10.1051/0004-6361/202346717}{\emph{Astron. Astrophys.}
  {\bfseries 678} (2023) A165}
  [\href{https://arxiv.org/abs/2304.10431}{{\ttfamily 2304.10431}}].

\bibitem{delabrouillePrelaunchPlanckSky2013}
J.~Delabrouille, M.~Betoule, J.-B.~Melin, M.-A.~{Miville-Desch{\^e}nes},
  J.~{Gonzalez-Nuevo}, M.L.~Jeune et~al., \emph{The pre-launch {{Planck Sky
  Model}}: A model of sky emission at submillimetre to centimetre wavelengths},
  \href{https://doi.org/10.1051/0004-6361/201220019}{\emph{Astron. Astrophys.}
  {\bfseries 553} (2013) A96}
  [\href{https://arxiv.org/abs/1207.3675}{{\ttfamily 1207.3675}}].

\bibitem{planckcollaborationPlanck2018Results2020a}
P.~Collaboration, Y.~Akrami, M.~Ashdown, J.~Aumont, C.~Baccigalupi,
  M.~Ballardini et~al., \emph{Planck 2018 results. {{IV}}. {{Diffuse}}
  component separation},
  \href{https://doi.org/10.1051/0004-6361/201833881}{\emph{Astron. Astrophys.}
  {\bfseries 641} (2020) A4}
  [\href{https://arxiv.org/abs/1807.06208}{{\ttfamily 1807.06208}}].

\bibitem{planckcollaborationPlanckIntermediateResults2016}
P.~Collaboration, N.~Aghanim, M.~Ashdown, J.~Aumont, C.~Baccigalupi,
  M.~Ballardini et~al., \emph{Planck intermediate results. {{XLVIII}}.
  {{Disentangling Galactic}} dust emission and cosmic infrared background
  anisotropies},
  \href{https://doi.org/10.1051/0004-6361/201629022}{\emph{Astron. Astrophys.}
  {\bfseries 596} (2016) A109}
  [\href{https://arxiv.org/abs/1605.09387}{{\ttfamily 1605.09387}}].

\bibitem{diego-palazuelosCosmicBirefringencePlanck2022}
P.~{Diego-Palazuelos}, J.R.~Eskilt, Y.~Minami, M.~Tristram, R.M.~Sullivan,
  A.J.~Banday et~al., \emph{Cosmic {{Birefringence}} from {{Planck Data
  Release}} 4},
  \href{https://doi.org/10.1103/PhysRevLett.128.091302}{\emph{Phys. Rev. Lett.}
  {\bfseries 128} (2022) 091302}
  [\href{https://arxiv.org/abs/2201.07682}{{\ttfamily 2201.07682}}].

\bibitem{thornePythonSkyModel2017}
B.~Thorne, J.~Dunkley, D.~Alonso and S.~Naess, \emph{The {{Python Sky Model}}:
  Software for simulating the {{Galactic}} microwave sky},
  \href{https://doi.org/10.1093/mnras/stx949}{\emph{Mon. Not. R. Astron. Soc.}
  {\bfseries 469} (2017) 2821}
  [\href{https://arxiv.org/abs/1608.02841}{{\ttfamily 1608.02841}}].

\bibitem{zoncaPythonSkyModel2021}
A.~Zonca, B.~Thorne, N.~Krachmalnicoff and J.~Borrill, \emph{The {{Python Sky
  Model}} 3 software}, \href{https://doi.org/10.21105/joss.03783}{\emph{JOSS}
  {\bfseries 6} (2021) 3783}
  [\href{https://arxiv.org/abs/2108.01444}{{\ttfamily 2108.01444}}].

\bibitem{liuMethodsPixelDomain2019}
H.~Liu, J.~Creswell, S.~{von Hausegger} and P.~Naselsky, \emph{Methods for
  pixel domain correction of {{EB}} leakage},
  \href{https://doi.org/10.1103/PhysRevD.100.023538}{\emph{Phys. Rev. D}
  {\bfseries 100} (2019) 023538}
  [\href{https://arxiv.org/abs/1811.04691}{{\ttfamily 1811.04691}}].

\bibitem{douForegroundRemovalILC2023}
J.~Dou, S.~Ghosh, L.~Santos and W.~Zhao, \emph{Foreground removal with {{ILC}}
  methods for {{AliCPT-1}}},  Oct., 2023.
\newblock 10.48550/arXiv.2310.19627.

\bibitem{alonsoUnifiedPseudoC_2019}
D.~Alonso, J.~Sanchez and A.~Slosar, \emph{A unified
  pseudo-\${{C}}\_{\textbackslash}ell\$ framework},
  \href{https://doi.org/10.1093/mnras/stz093}{\emph{Mon. Not. R. Astron. Soc.}
  {\bfseries 484} (2019) 4127}
  [\href{https://arxiv.org/abs/1809.09603}{{\ttfamily 1809.09603}}].

\bibitem{aumontAbsoluteCalibrationPolarisation2020}
J.~Aumont, J.F.~{Mac{\'i}as-P{\'e}rez}, A.~Ritacco, N.~Ponthieu and
  A.~Mangilli, \emph{Absolute calibration of the polarisation angle for future
  {{CMB}} {{B}}-mode experiments from current and future measurements of the
  {{Crab}} nebula},
  \href{https://doi.org/10.1051/0004-6361/201833504}{\emph{Astron. Astrophys.}
  {\bfseries 634} (2020) A100}
  [\href{https://arxiv.org/abs/1805.10475}{{\ttfamily 1805.10475}}].

\bibitem{minamiNewExtractionCosmic2020}
Y.~Minami and E.~Komatsu, \emph{New {{Extraction}} of the {{Cosmic
  Birefringence}} from the {{Planck}} 2018 {{Polarization Data}}},
  \href{https://doi.org/10.1103/PhysRevLett.125.221301}{\emph{Phys. Rev. Lett.}
  {\bfseries 125} (2020) 221301}.

\bibitem{eskiltCosmoglobeDR1Results2023}
J.R.~Eskilt, D.J.~Watts, R.~Aurlien, A.~Basyrov, M.~Bersanelli, M.~Brilenkov
  et~al., \emph{Cosmoglobe {{DR1}} results. {{II}}. {{Constraints}} on
  isotropic cosmic birefringence from reprocessed {{WMAP}} and {{Planck LFI}}
  data}, \href{https://doi.org/10.1051/0004-6361/202346829}{\emph{Astron.
  Astrophys.} {\bfseries 679} (2023) A144}
  [\href{https://arxiv.org/abs/2305.02268}{{\ttfamily 2305.02268}}].

\bibitem{eriksenForegroundRemovalWilkinson2004}
H.K.~Eriksen, A.J.~Banday, K.M.~Gorski and P.B.~Lilje, \emph{On {{Foreground
  Removal}} from the {{Wilkinson Microwave Anisotropy Probe Data}} by an
  {{Internal Linear Combination Method}}: {{Limitations}} and
  {{Implications}}}, \href{https://doi.org/10.1086/422807}{\emph{ApJ}
  {\bfseries 612} (2004) 633}
  [\href{https://arxiv.org/abs/astro-ph/0403098}{{\ttfamily
  astro-ph/0403098}}].

\bibitem{tegmarkHighResolutionForeground2003}
M.~Tegmark, A.~{de Oliveira-Costa} and A.~Hamilton, \emph{A high resolution
  foreground cleaned {{CMB}} map from {{WMAP}}},
  \href{https://doi.org/10.1103/PhysRevD.68.123523}{\emph{Phys. Rev. D}
  {\bfseries 68} (2003) 123523}
  [\href{https://arxiv.org/abs/astro-ph/0302496}{{\ttfamily
  astro-ph/0302496}}].

\bibitem{basakNeedletILCAnalysis2012}
S.~Basak and J.~Delabrouille, \emph{A needlet {{ILC}} analysis of {{WMAP}}
  7-year data: Estimation of {{CMB}} temperature map and power spectrum},
  \href{https://doi.org/10.1111/j.1365-2966.2011.19770.x}{\emph{Mon. Not. R.
  Astron. Soc.} {\bfseries 419} (2012) 1163}
  [\href{https://arxiv.org/abs/1106.5383}{{\ttfamily 1106.5383}}].

\bibitem{basakNeedletILCAnalysis2013}
S.~Basak and J.~Delabrouille, \emph{A needlet {{ILC}} analysis of {{WMAP}}
  9-year polarisation data: {{CMB}} polarisation power spectra},
  \href{https://doi.org/10.1093/mnras/stt1158}{\emph{Monthly Notices of the
  Royal Astronomical Society} {\bfseries 435} (2013) 18}
  [\href{https://arxiv.org/abs/1204.0292}{{\ttfamily 1204.0292}}].

\bibitem{remazeillesExploringCosmicOrigins2017}
M.~Remazeilles, A.J.~Banday, C.~Baccigalupi, S.~Basak, A.~Bonaldi, G.~De~Zotti
  et~al., \emph{Exploring {{Cosmic Origins}} with {{CORE}}: {{B-mode Component
  Separation}}},  \href{https://arxiv.org/abs/1704.04501}{{\ttfamily
  1704.04501}}.

\bibitem{adakBModeForecastCMBBh2021}
D.~Adak, A.~Sen, S.~Basak, J.~Delabrouille, T.~Ghosh, A.~Rotti et~al.,
  \emph{B-mode forecast of {{CMB-Bh}}$\overline a$rat},
  \href{https://arxiv.org/abs/2110.12362}{{\ttfamily 2110.12362}}.

\bibitem{krachmalnicoffInflightPolarizationAngle2022}
N.~Krachmalnicoff, T.~Matsumura, E.~{de la Hoz}, S.~Basak, A.~Gruppuso,
  Y.~Minami et~al., \emph{In-flight polarization angle calibration for
  {{LiteBIRD}}: Blind challenge and cosmological implications},
  \href{https://doi.org/10.1088/1475-7516/2022/01/039}{\emph{J. Cosmol.
  Astropart. Phys.} {\bfseries 2022} (2022) 039}
  [\href{https://arxiv.org/abs/2111.09140}{{\ttfamily 2111.09140}}].

\bibitem{zhangEfficientILCAnalysis2022}
Z.~Zhang, Y.~Liu, S.-Y.~Li, D.-L.~Wu, H.~Li and H.~Li, \emph{Efficient {{ILC}}
  analysis on polarization maps after {{EB}} leakage correction},
  \href{https://doi.org/10.1088/1475-7516/2022/07/044}{\emph{J. Cosmol.
  Astropart. Phys.} {\bfseries 2022} (2022) 044}
  [\href{https://arxiv.org/abs/2109.12619}{{\ttfamily 2109.12619}}].

\bibitem{caronesAnalysisNILCPerformance2023}
A.~Carones, M.~Migliaccio, D.~Marinucci and N.~Vittorio, \emph{Analysis of
  {{NILC}} performance on {{B-modes}} data of sub-orbital experiments},
  \href{https://doi.org/10.1051/0004-6361/202244824}{\emph{Astron. Astrophys.}
  {\bfseries 677} (2023) A147}
  [\href{https://arxiv.org/abs/2208.12059}{{\ttfamily 2208.12059}}].

\bibitem{aurlienForegroundSeparationConstraints2023}
R.~Aurlien, M.~Remazeilles, S.~Belkner, J.~Carron, J.~Delabrouille,
  H.K.~Eriksen et~al., \emph{Foreground {{Separation}} and {{Constraints}} on
  {{Primordial Gravitational Waves}} with the {{PICO Space Mission}}},
  \href{https://doi.org/10.1088/1475-7516/2023/06/034}{\emph{JCAP} {\bfseries
  06} (2023) 034} [\href{https://arxiv.org/abs/2211.14342}{{\ttfamily
  2211.14342}}].

\bibitem{fuskelandTensortoscalarRatioForecasts2023}
U.~Fuskeland, J.~Aumont, R.~Aurlien, C.~Baccigalupi, A.J.~Banday, H.K.~Eriksen
  et~al., \emph{Tensor-to-scalar ratio forecasts for extended {{LiteBIRD}}
  frequency configurations},
  \href{https://arxiv.org/abs/2302.05228}{{\ttfamily 2302.05228}}.

\bibitem{planckcollaborationPlanckIntermediateResults2016a}
P.~Collaboration, R.~Adam, P.A.R.~Ade, N.~Aghanim, M.~Arnaud, J.~Aumont et~al.,
  \emph{Planck intermediate results. {{XXX}}. {{The}} angular power spectrum of
  polarized dust emission at intermediate and high {{Galactic}} latitudes},
  \href{https://doi.org/10.1051/0004-6361/201425034}{\emph{Astron. Astrophys.}
  {\bfseries 586} (2016) A133}
  [\href{https://arxiv.org/abs/1409.5738}{{\ttfamily 1409.5738}}].

\bibitem{zhaoSeparatingTypesPolarization2010}
W.~Zhao and D.~Baskaran, \emph{Separating {{E}} and {{B}} types of polarization
  on an incomplete sky},
  \href{https://doi.org/10.1103/PhysRevD.82.023001}{\emph{Phys. Rev. D}
  {\bfseries 82} (2010) 023001}
  [\href{https://arxiv.org/abs/1005.1201}{{\ttfamily 1005.1201}}].

\bibitem{ghoshEndingPartialSky2021}
S.~Ghosh, J.~Delabrouille, W.~Zhao and L.~Santos, \emph{Towards ending the
  partial sky {{E-B}} ambiguity in {{CMB}} observations},
  \href{https://doi.org/10.1088/1475-7516/2021/02/036}{\emph{J. Cosmol.
  Astropart. Phys.} {\bfseries 2021} (2021) 036}
  [\href{https://arxiv.org/abs/2007.09928}{{\ttfamily 2007.09928}}].

\end{thebibliography}\endgroup
\bibliographystyle{JHEP}

\end{document}